\documentclass[11pt,english]{article}

\usepackage[latin9]{inputenc}
\usepackage{geometry}
\usepackage{amsmath}
\usepackage[authoryear]{natbib}
\usepackage[para,online,flushleft]{threeparttable} 

\usepackage{natbib}
\usepackage{mathpazo} 
\usepackage[T1]{fontenc}

\usepackage{array}
\newcolumntype{P}[1]{>{\raggedright\arraybackslash}p{#1}}
\newcolumntype{D}[1]{>{\raggedleft\arraybackslash}p{#1}}

\usepackage{changepage} 
\usepackage[dvipsnames]{xcolor}
\usepackage{multirow}
\usepackage{graphics}
\usepackage{graphicx}
\usepackage{subfigure}
\usepackage{siunitx}
\usepackage{babel}
\usepackage{dcolumn,booktabs}
\usepackage[capposition=bottom]{floatrow}

\usepackage{setspace} 

\def\sym#1{\ifmmode^{#1}\else\(^{#1}\)\fi} 

\usepackage{hyperref}
\definecolor{myred}{RGB}{100,0,0} 

\hypersetup{colorlinks=true,linkcolor=myred,citecolor=myred}

\newenvironment{figurenotes}[1][Note]{\begin{minipage}[t]{\linewidth}\footnotesize{\itshape#1: }}{\end{minipage}}

\newenvironment{ldcnotes}{\begin{minipage}[t]{\linewidth}\footnotesize{ }}{\end{minipage}}

\begin{document}

\begin{titlepage}
\title{Dynamic Beveridge Curve Accounting\thanks{We thank Katharine Abraham, Andrew Figura, David Ratner, and seminar participants at the 2019 SOLE meeting, the Fall 2018 Midwest Macro meeting, and the Federal Reserve Board. Vivi Gregorich provided excellent research assistance. Opinions expressed herein are those of the authors alone and do not necessarily reflect the views of the Federal Reserve System or the Board of Governors.}}
\author{Hie Joo Ahn \and Leland D. Crane\thanks{Ahn: Federal Reserve Board of Governors, hiejoo.ahn@frb.gov.  Crane: Federal Reserve Board of Governors, leland.d.crane@frb.gov.}}
\date{\today}
\maketitle
\begin{abstract}
\noindent We develop a dynamic decomposition of the empirical Beveridge curve, i.e., the level of vacancies conditional on unemployment. Using a standard model, we show that three factors can shift the Beveridge curve: reduced-form matching efficiency, changes in the job separation rate, and out-of-steady-state dynamics. We find that the shift in the Beveridge curve during and after the Great Recession was due to all three factors, and each factor taken separately had a large effect. Comparing the pre-2010 period to the post-2010 period, a fall in matching efficiency and out-of-steady-state dynamics both pushed the curve upward, while the changes in the separation rate pushed the curve downward. The net effect was the observed upward shift in vacancies given unemployment. In previous recessions changes in matching efficiency were relatively unimportant, while dynamics and the separation rate had more impact. Thus, the unusual feature of the Great Recession was the deterioration in matching efficiency, while separations and dynamics have played significant, partially offsetting roles in most downturns. The importance of these latter two margins contrasts with much of the literature, which abstracts from one or both of them.  We show that these factors affect the slope of the empirical Beveridge curve, an important quantity in recent welfare analyses estimating the natural rate of unemployment. 

\end{abstract}
\thispagestyle{empty}

\end{titlepage}

\setcounter{page}{1}
\section{Introduction}

\begin{figure}[h]
\includegraphics[width=0.7\textwidth]{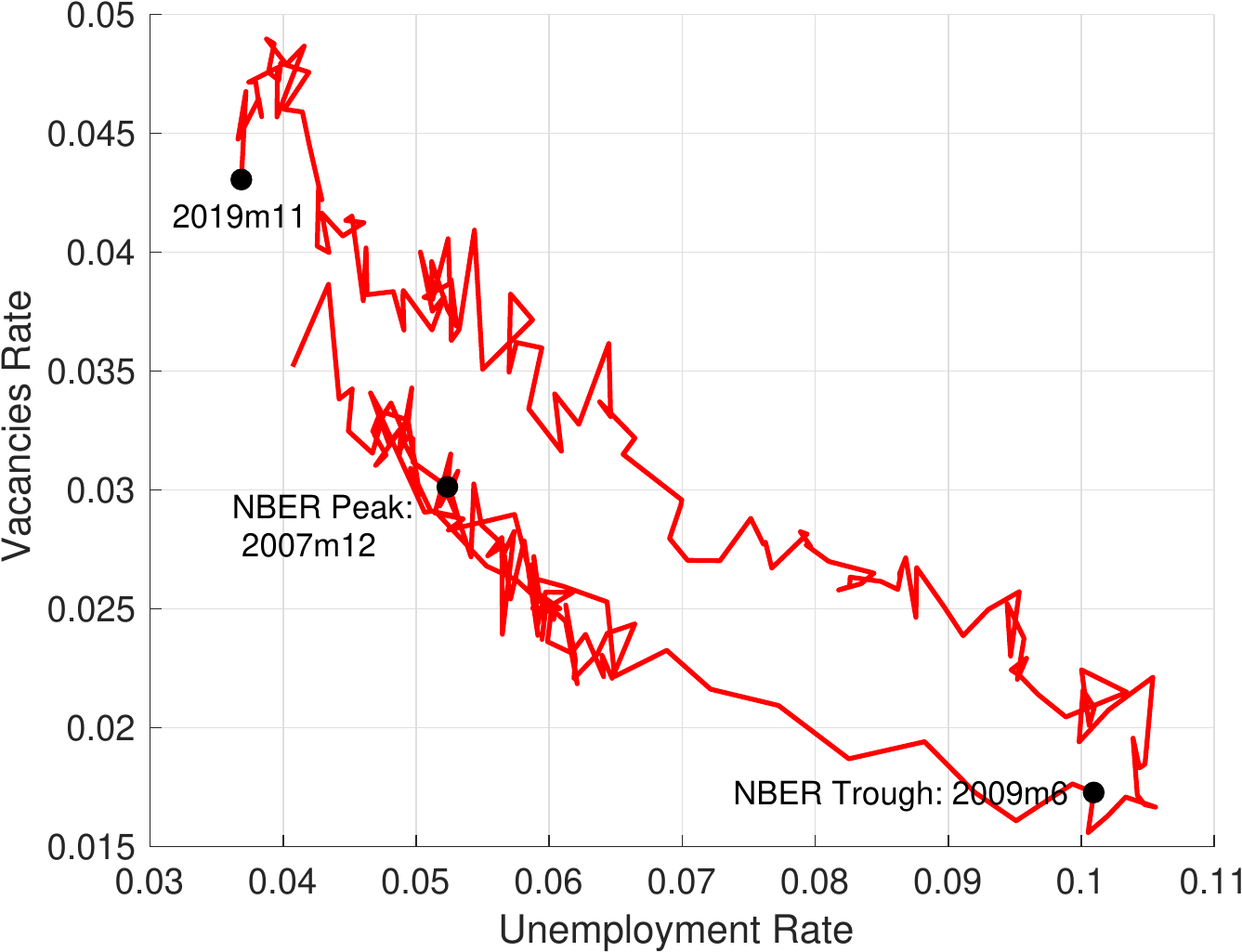}
\caption{The Beveridge curve}
\label{fig:basic_BC}
\begin{ldcnotes}
{\itshape Note:} Monthly data, 2000-2019. 
\newline
{\itshape Source:} CPS, JOLTS. 
\end{ldcnotes}
\end{figure}

The empirical Beveridge curve\textemdash the level of vacancies conditional on unemployment\textemdash has long been of interest to economists and policy makers.  Interest intensified in the wake of the Great Recession, as the curve appeared to shift upwards (see Figure \ref{fig:basic_BC}), fueling concerns about the functioning of the labor market.   There is not currently consensus on the cause of this shift (or historical Beveridge curve shifts).  Many papers have attributed the shift to falling matching efficiency (whether due to mismatch, duration dependence, recruiting intensity, heterogeneity, or other causes.)  Others researchers have argued that mechanical out-of-steady state dynamics can account for the apparent shift.  Finally, it has also been noted that variation in the employment separation rate can also produce shifts in the Beveridge curve.  Each of these threads of the literature has taken a slightly different modelling approach as, some authors use steady-state approximations, while others assume a constant job separation rate.  

In this paper we provide a new, unified accounting model for the Beveridge curve and a related decomposition method. In our baseline model, where the labor-force status is either employed or unemployed, there are three main factors that matter for the position of the Beveridge curve: (1) matching efficiency, (2) the job-separation probability, and (3) out-of-steady-state dynamics. We analyze how much each of these factors shifted the Beveridge curve. The model allows us to estimate how the contribution of each factor changed in different recessionary and recovery episodes. We also extend our model to include the labor-force participation margin, to see how important labor-supply factors are in the dynamics of Beveridge curve. 

We find that matching efficiency, job separations and out-of-steady-state dynamics are \emph{all} important in understanding the shifts of the Beveridge curve over business cycles, particularly in the Great Recession.  Out-of-steady state dynamics (defined below) produced a net upward shift in the Beveridge curve during and after the Great Recession, as suggested by \cite{CET2015} and \cite{furlanetto2016}.\footnote{See also \cite{eichenbaum2015} for related discussion.}  Those papers assume a constant job separation rates, but we find that changes in the job separation rate shifted the Beveridge curve sharply \emph{down} on net around the Great Recession. This downward shift of the Beveridge curve partially offset the combined upward shift from out-of-steady-state dynamics and matching efficiency.  In fact, changes in the separation rate were the largest single factor moving the Beveridge curve.  Separations can shift the Beveridge curve since, for a given path of unemployment, a higher separation rate implies that vacancies must also be higher,  in order to  maintain the net change in unemployment at the observed values.  The job separation probability was high in the downswing of the Great Recession, and it later fell back to more normal levels in the recovery.  This had the effect of shifting the Beveridge curve up in the downswing and down in upswing. \cite{elsby2015} documented a similar point, though they did not quantify the extent of the shift or compare it to the other shifters.\footnote{\cite{hall_wohl2018} also noted the unemployment inflow rate complicates the behavior of the Beveridge curve.}   We also find that matching efficiency fell significantly during and after the Great Recession, which pushed the Beveridge curve up.  This result is consistent with, e.g., \cite{barnichon2015}.\footnote{See also \cite{barnichon2010b} and \cite{barnichon2012} for more on matching efficiency.}  

Analyses which ignore one or more of these shifters will either fail to match the data or will risk making mistaken inferences.  This leads to several concrete conclusions and recommendations: First, the importance of out-of-steady-state dynamics implies that the usual flow steady-state approximations are \emph{not} appropriate for studying the Great Recession, or similar periods of rapid change in the unemployment rate.  Flow steady-state approximations have become a fundamental tool for simplifying and understanding the labor market (see, for example, \cite{fujita2009},  \cite{elsby2009}, \cite{shimer2012}, \cite{barnichon2012},  \cite{elsby2015}.)  Unfortunately, in the Great Recession unemployment was consistently far from the the steady steady-state value implied by inflows and outflows, thus the approximation is poor during this period.  We also find a large role for out-of-steady-state dynamics in some previous recessions.

Second, time-variation in the job separation probability is critical for understanding the Beveridge curve, and indeed was the single largest shifter of the Beveridge curve in the Great Recession.  Thus, the common simplifying assumption of a constant separation rate (made in, e.g., \cite{CET2015}) is not appropriate when trying to model the Beveridge curve.  In fact, we find that variation in the separation rate was an important shifter of the Beveridge curve in many previous recessions as well, and this variation also affects the slope of the empirical curve.  Our analysis \emph{does not} speak directly to the debate over the relative importance of the separations versus the job findings for the evolution of unemployment (see, e.g.,  \cite{fujita2009},  \cite{elsby2009}, \cite{shimer2012}, \cite{ahn2016}).  Rather, we simply point out that the Beveridge curve cannot be properly understood without this ingredient.

Third, we confirm that there was a clear fall in reduced-form matching efficiency in the Great Recession, as has been documented in several other papers (see \cite{elsby2010}, \cite{barnichon2015}).  We show that this drop in matching efficiency shifted the Beveridge curve substantially and persistently upward in the Great Recession (though the other shifters partially obscure this effect.)  In this paper we do not attempt to explain \emph{why} matching efficiency fell, instead we seek to quantify the effects on the Beveridge curve and the interactions with other factors.\footnote{Many papers have offered explanations for the fall in reduced-form matching efficiency among them \cite{davis2013}, \cite{sahin2014}, \cite{elsby2015}, \cite{barnichon2015}, \cite{kroft2015}, \cite{ahn2016},  and \cite{hall_wohl2018}.}

Though all three of these factors are crucial in understanding the Beveridge curve, we also find that the relative importance of each factor differed across  recessionary episodes.\footnote{\cite{daly2011} and \cite{diamond2015} document historical Beveridge curve shifts.}   We find that the 1990's recession was similar to the Great Recession in that matching efficiency was the key factor to the persistent outward shift of Beveridge curve. However, in the other recessions in the 1970's, 1980's and 2001, the job separation probability and out-of-steady-state dynamics played more important roles than matching efficiency. 

In addition to clarifying the source of loops in the Beveridge curve, we show that these shifters affect the \emph{slope} of the empirical Beveridge curve.  This occurs because the curve is being shifted while labor market upswings and downswings progress, not just at peaks and troughs.  Thus the slope of the steady-state Beveridge curve under constant separations and constant matching efficiency is very different from the empirical slope.  This has direct implications for the work of \cite{michaillat2019}, who exploit the slope of the Beveridge curve to estimate the efficient level of unemployment and the unemployment gap.  A back of the envelope exercise shows that using an arguably more appropriate slope cuts the estimated unemployment gap in half, relative to \cite{michaillat2019}.  We view this as evidence that more work is needed to understand how time-varying factors affect the slope of the empirical Beveridge curve.

For our baseline results, we work with a log-linearized Beveridge curve, which expresses the vacancy rate a linear function of various factors.  This first-order approximation matches the observed Beveridge curve quite well, and the factors and their associated coefficients are easily interpretable.  This analytical tool makes it easy to trace out the contributions of factors to the shifts in the Beveridge curve, and trace out counterfactual curves that hold various factors constant.

In might be worried that results based on a Taylor series approximation can be inaccurate.  In addition, under an approximate Beveridge curve the implied paths of vacancies will not be exactly consistent with the matching function and the law of motion for unemployment.  To address this concern we perform similar decompositions, holding various factors constant, using the actual, non-linear Beveridge curve relation, and show that the results are nearly unchanged.  Of course, when using the non-linear version the exact contributions of each margin depend on the ordering of the variables in the decomposition.  But the results are qualitatively consistent across all orderings.

The next section introduces the basic model.  Section \ref{sec:data} discusses the data.  Section \ref{sec:linearization} linearizes the model and presents the results for the Great Recession.    Historical recessions are covered in Section \ref{sec:historical_recessions}, and the results of a three-state model are discussed in Section \ref{sec:3state}.  Section \ref{sec:conclusion} concludes. Appendix \ref{sec:full_decompositions} addresses the robustness of the linearized results by calculating exact non-linear decompositions.

\section{Model}\label{sec:model}

This section derives a version of the simple Beveridge curve framework used in \cite{CET2015} (hereafter CET) and \cite{eichenbaum2015}, which is nearly identical to that of \cite{elsby2015}.  We do not close the model by making assumptions about the job creation process, wage determination, or other fundamentals.  Instead we focus on deriving conclusions that must hold for any general equilibrium model whose labor market is described by (1) the standard law of motion for unemployment and (2) the usual matching function relationship.
 
Let $U_{t}$ be the unemployment rate in month $t$, and let $V_{t}$ be the vacancy rate (i.e. vacancies divided by the labor force).  There is no on-the-job search, no participation margin, and the size of the labor force is constant and normalized to unity.

\begin{equation}
H_{t}=\sigma_{t} U_{t}^{1-\alpha}V_{t}^{\alpha}\label{eq:hires}
\end{equation}

\noindent where $\alpha$ is the elasticity of the matching function and $\sigma_{t}$ is matching efficiency, which can vary over time.  Then the job-finding probability is given by
\begin{equation}
f_{t}=\sigma_{t}(V_{t}/U_{t})^{\alpha}.\label{eq:jf_rate}
\end{equation}
The law of motion for unemployment is 

\begin{equation}
U_{t+1}=s_{t}\left(1-U_{t}\right)-f_{t}U_{t}+U_{t}\label{eq:U_LOM}
\end{equation}
where $s_{t}$ is the probability a job ends in a given month.  We refer to $s_{t}$ as the ``EU probability'', as it is the probability an employed worker transitions to unemployment in a given month.  Substituting equation \eqref{eq:jf_rate} into \eqref{eq:U_LOM} and rearranging we arrive at
\begin{equation}
V_{t}=\left[\frac{s_{t}(1-U_{t})- \Delta U_{t+1} }{\sigma_{t}U_{t}^{1-\alpha}}\right]^{1/\alpha}\label{eq:CET_identity}
\end{equation}
\noindent where $\Delta U_{t+1}=U_{t+1}-U_{t}$.  This is a slight generalization of CET equation 5.2. Whereas CET assume that $s_{t}$ and $\sigma_{t}$ are constants, we permit time-variation in these parameters.  Note that if $s_{t}$ is set to its observed values and $\sigma_{t}$ is chosen to verify equation \eqref{eq:hires}, then equation \eqref{eq:CET_identity} is an identity. 

Equation \eqref{eq:CET_identity} is at the core of our analysis.  To understand it better, consider the case where $s_{t}$, $\sigma_{t}$ and $U_{t}$ are constants:
\begin{equation}
V=\left[ \frac{ s (1-U) }{\sigma U^{1-\alpha}}\right]^{1/\alpha}\label{eq:SSBC}.
\end{equation} 
\noindent This is the steady state Beveridge curve relationship at the core of textbook search models (see \cite{pissarides2000}):  a steady-state with low $U$ must have high $V$, and vice-versa.  Taking  equation \eqref{eq:SSBC} as the reference point, variation in $s_{t}$, $\sigma_{t}$ and $\Delta U_{t+1}$ changes the level of $V_{t}$ given $U_{t}$.  Thus, with a slight abuse of terminology, we will refer to these factors as \emph{shifters}.\footnote{We use shifters to mean factors that change $V_{t}$ given $U_{t}$. Note $s$ and $\sigma$ also shift the steady-state Beveridge curve \eqref{eq:SSBC}, while $\Delta U_{t+1}$ does not.  The dynamics captured by $\Delta U_{t+1}$ produce loops \emph{around} the steady-state Beveridge curve, but do not change that model-based relationship.  }

Given a path for unemployment and hypothesized, possibly counterfactual, values of the parameters $\left(\alpha,s_{t},\sigma_{t}\right)$, one can calculate the implied path of vacancies from equation \eqref{eq:CET_identity} and compare it to the true path of vacancies.  This is the essence of our exercises in Section \ref{sec:linearization}.  


\section{Data}\label{sec:data}

\begin{figure}[htbp]
\includegraphics[width=0.6\textwidth]{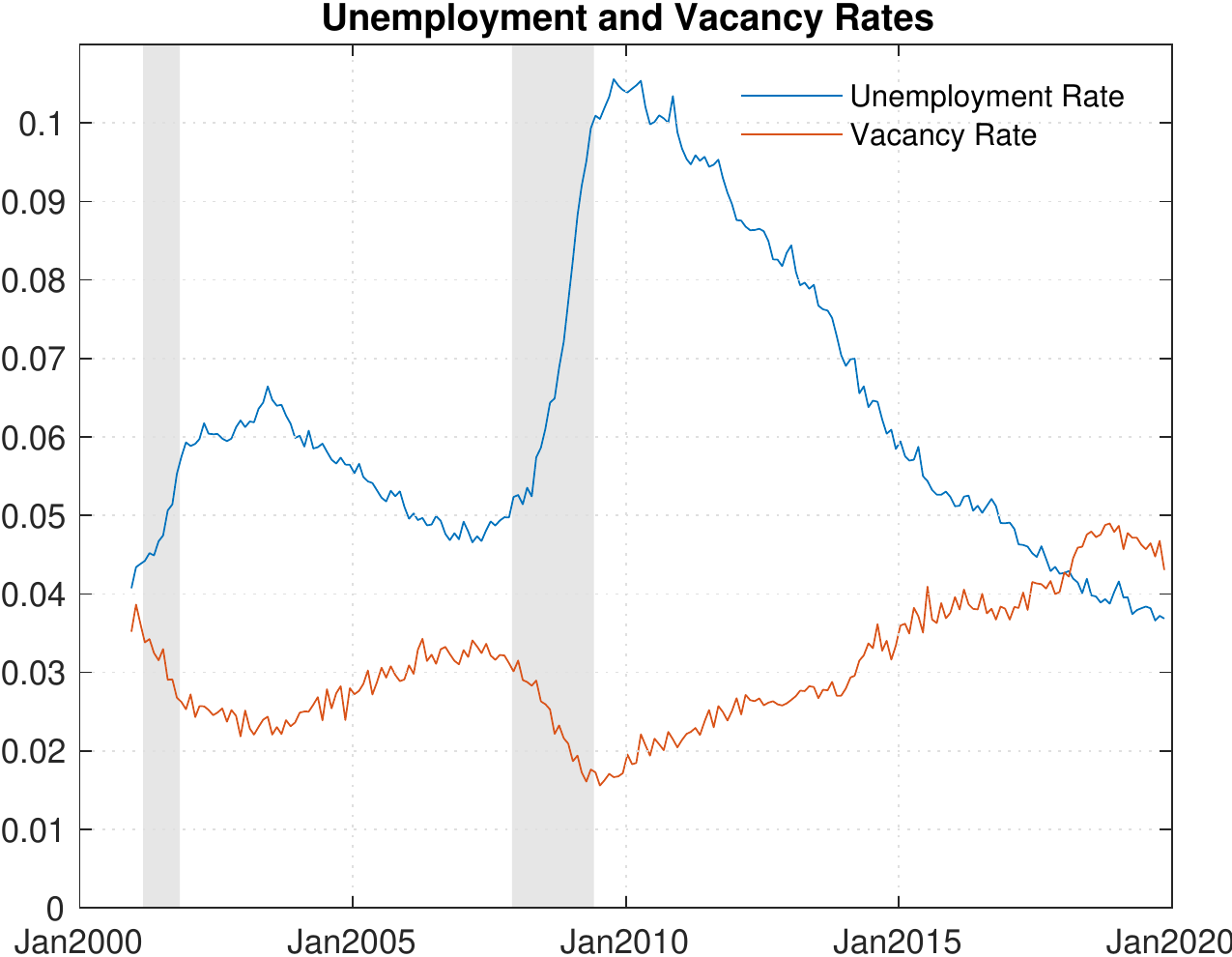}
\caption{Unemployment and Vacancy Rates}
\label{fig:U_and_V}
\begin{ldcnotes}
{\itshape Note:} Monthly data, 2000-2019. NBER recessions shaded in gray.
\newline
{\itshape Source:} CPS, JOLTS. 
\end{ldcnotes}
\end{figure}

We require data on all the variables and parameters in equations \eqref{eq:U_LOM} and \eqref{eq:CET_identity}.  We use the standard approaches, based mostly on \cite{shimer2012} and \cite{barnichon2015}.  We set $U_{t}$ as the number of unemployed divided by the labor force, as measured in the Current Population Survey (CPS).  We set $V_{t}$ equal to the count of vacancies from JOLTS divided by the size of the labor force.  Figure \ref{fig:U_and_V} plots the two series.   

We set the monthly job-finding probability, $f_{t}$ as in \cite{shimer2012}, using data on the number of short-term unemployed each month.\footnote{That is, we set $f_{t}=1-\frac{U_{t+1}-U^{s}_{t+1}}{U_{t}}$, where $U^{s}_{t+1}$ is the number of workers unemployed for less than five weeks in month $t+1$.  Thus $f_{t}$ is the probability that a worker unemployed in month $t$ finds a job by $t+1$.  In the data it is possible for such a worker to both find and lose a job (or multiple jobs) before $t+1$, but the discrete-time model we use rules out this possibility.} We then choose $s_{t}$ to satisfy the law of motion \eqref{eq:U_LOM} exactly.\footnote{In both our setup and the continuous time formulation of \cite{shimer2012}, EU flows are set so as to make the observed sequence of stocks consistent with the flows.  In the three-state model of Section \ref{sec:3state} the transition rates are taken directly from the data and raked for consistency with the stocks.}    

\begin{figure}[htbp]
\includegraphics[width=0.6\textwidth]{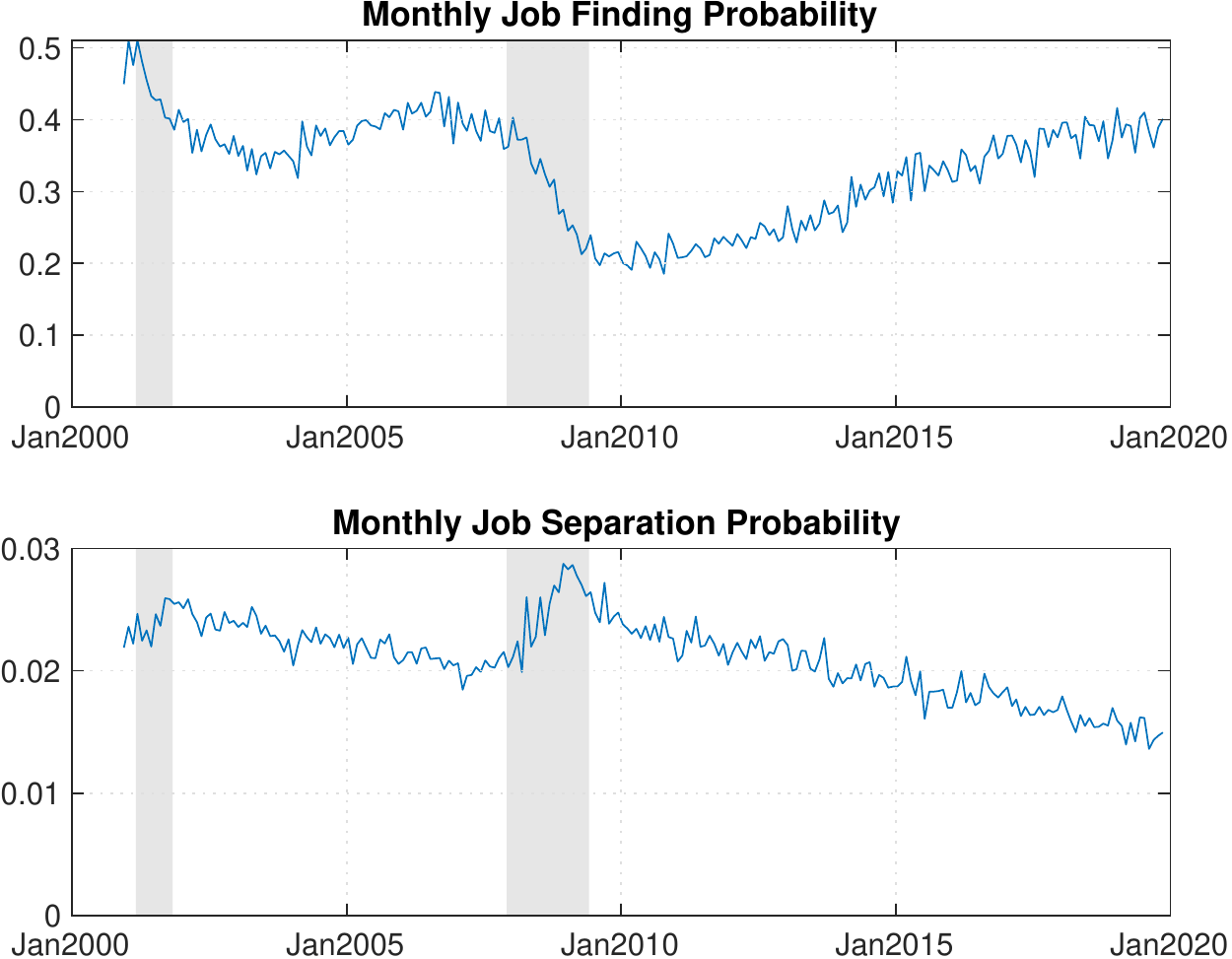}
\caption{Observed Transition Probabilities}
\label{fig:jf_sep_data}
\begin{ldcnotes}
{\itshape Note:} Monthly data, 2000-2019. NBER recessions shaded in gray.
\newline
{\itshape Source:} CPS, JOLTS. 
\end{ldcnotes}
\end{figure}

Figure \ref{fig:jf_sep_data} shows the job finding and separation probabilities.  It is notable that the job finding probability fell by about 50 percent in the Great Recession and the separation probability increased by about 50 percent.\footnote{\cite{CET2015} note that the job separation rate, as measured by JOLTS, \emph{fell} in the Great Recession.  The JOLTS separation rate includes job-to-job flows, which are known to be highly procylical, as well as flows to nonemployment.  Their model, like ours, does not allow for job-to-job flows.  The JOLTS separation rate is likely the correct measure when considering the firm's problem, since it gives the expected duration of the match.  But when considering the evolution of unemployment it is better to use the inflow to unemployment, rather than including job-to-job flows.}  This suggests that both margins may have played a significant role in the evolution of unemployment.  We will confirm this impression in what follows. 

Measurement of $\alpha$ and $\sigma_{t}$ require estimation of the matching function.  We run the usual regression 

\begin{equation}
\ln f_{t}=\ln \overline{\sigma}  +  \alpha \ln \left( \frac{V_{t}}{U_{t}} \right)  +\varepsilon_{t} \label{eq:matching_regress}
\end{equation}
where $\varepsilon_{t}$ is the mean-zero error term, $\sigma_{t}= \overline{\sigma} \exp (\varepsilon_{t})$ is time-varying matching efficiency, and $\overline{\sigma}$ is interpreted as average matching efficiency.

\begin{figure}[htbp]
\includegraphics[width=0.6\textwidth]{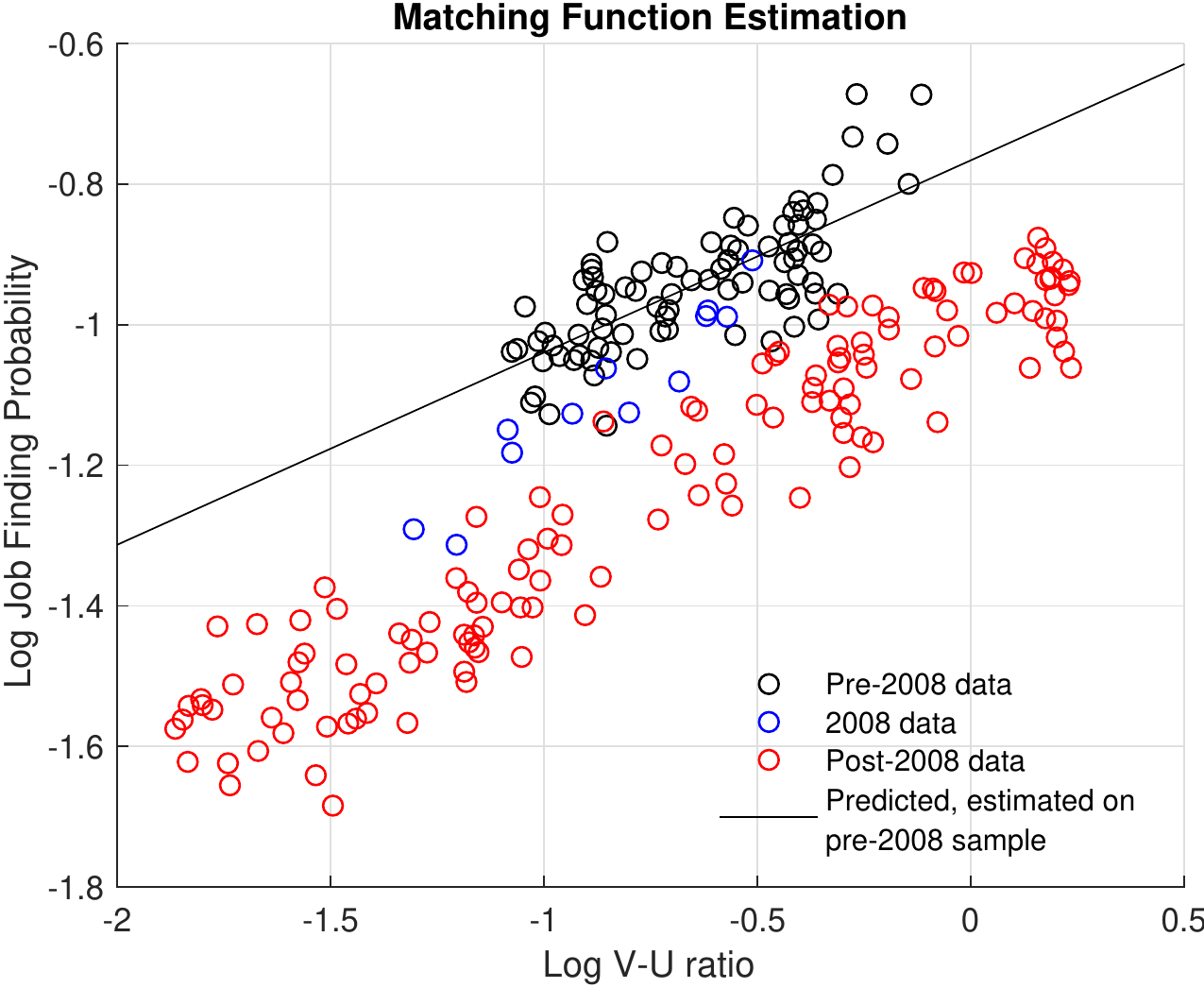}
\caption{Matching Function Estimation}
\label{fig:match_fun}
\begin{figurenotes}
Monthly data, 2000-2019. 
\end{figurenotes}
\begin{figurenotes}[Source]
CPS, JOLTS.
\end{figurenotes}
\end{figure}

Figure \ref{fig:match_fun} plots the log job finding probability against the log V-U ratio.  The data for different periods are plotted in different colors.  It is evident that matching efficiency deteriorated significantly post-2008.  Any change in the matching elasticity $\alpha$ was minor by comparison, so we will continue assuming that $\alpha$ is a constant throughout the paper (as is standard in the literature).  

We run equation \eqref{eq:matching_regress} on a sample starting in 2000 (when the JOLTS series begins) and ending in 2007, a period where it is plausible that $\sigma$ was indeed constant.  We also run the regression on a post-2008 sample.  Table \ref{tab:match_fun_table} presents the results. The point estimates put $\alpha$ near 0.3, very similar to the estimates of  \cite{shimer2005} and \cite{barnichon2015}, who use longer time series.  It is evident that average matching efficiency fell about 25\% between the two samples.  

\begin{table}
    \centering
    \caption{Matching Function Estimates}
    \label{tab:match_fun_table}
    \begin{threeparttable}
    \sisetup{                                                 
  input-symbols=(),                                       
  table-format=-1.5,                                      
  table-space-text-post=***,                              
  table-align-text-post=false,                            
  group-digits=false                                      
}                                                         
\begin{tabular}{SSS}                                      
 & {(1)} & {(2)} \\                                       
 & {Pre-2008 Sample} & {Post-2008 Sample} \\ \hline \hline
{$\ln \overline{\sigma}$} & -0.77***  & -1.00*** \\       
                        & (0.02)  & (0.01) \\             
               {$\alpha$} & 0.27***  & 0.34*** \\         
                       & (0.03)  & (0.01)                 
\end{tabular}                                             
                                                              
      \begin{tablenotes}
    \item \footnotesize
    \textit{Notes}: OLS estimates of average matching efficiency ($\ln \overline{\sigma}$) and the matching function elasticity ($\alpha$).  *, **, and *** 
    indicate statistical significance at the 10\%, 5\%, and 1\% levels, respectively.  Standard errors are in parentheses.
    \end{tablenotes}
    \end{threeparttable}
\end{table}

\section{Linearization and Results}\label{sec:linearization}

In order to simplify the discussion, we log-linearize equation \eqref{eq:CET_identity}.
In particular, we take the first order Taylor approximation around a point $\left( U_{t},s_{t},\sigma_{t},\Delta U_{t+1} \right)=\left( \overline{U}, \overline{s}, \overline{\sigma},0 \right)$.  The result is the following expression 

\begin{align}
\ln V_{t}	\approx & \ln \overline{V} -\left(\frac{\overline{U}}{\alpha\left(1-\overline{U}\right)} +\frac{1-\alpha}{\alpha}\right)\left(\ln U_{t}-\ln\overline{U}\right) \nonumber \\
	& \underbrace{ -\frac{\overline{U}}{\alpha\overline{s}(1-\overline{U})}  \Delta \ln U_{t+1} }_{\substack{\textit{Shift due to} \\ \textit{Dynamics}}} 
    \underbrace{ +\frac{1}{\alpha(1-\overline{U})}\left(\ln s_{t}-\ln\overline{s}\right) }_{\substack{\textit{Shift due to} \\ \textit{Separations}}}  
    \underbrace{ -\frac{1}{\alpha}\left(\ln \sigma_{t}-\ln\overline{\sigma}\right) }_{\substack{\textit{Shift due to} \\ \textit{Matching Efficiency}}}  \label{eq:loglin}
\end{align}
where $\overline{V}$ is equation \eqref{eq:CET_identity} evaluated at $\left( \overline{U}, \overline{s}, \overline{\sigma},0 \right)$.  

The first line of equation  \eqref{eq:loglin}  is the (approximate) steady-state Beveridge curve.  The second line contains the ``shifters''.  Treating $\ln V_{t}$ as a linear function of $\ln U_{t}$, these shifters move the y-intercept of the steady-state curve up and down.  For example, we can see that when unemployment is rising ($\Delta \ln U_{t+1}>0$) then $\ln V_{t}$ will be lower than the steady state curve.  This is because, all else equal, rising unemployment implies low finding and thus low $\ln V_{t}$, which is the out-of-steady-state dynamics mechanism outlined in \cite{pissarides2000}.

While increasing in $\Delta \ln U_{t+1}$ shifts $\ln V_{t}$ down, increases in the job separation probability $s_{t}$ shift the curve up.  The intuition is that a higher job-separation probability, conditional on a fixed value of $\Delta \ln U_{t+1}$, requires more equilibrium vacancies to absorb the unemployment inflows.  Increases in matching efficiency $\sigma_{t}$ obviously shift the curve down, as fewer vacancies are needed to rationalize the observed value of $\Delta \ln U_{t+1}$.  

We are interested in approximating the Beveridge curve around the Great Recession.  To that end, we center the Taylor approximation around post-2007 averages.  This yields $\overline{U}=0.068$, $\overline{s}=0.020$, and $\overline{\sigma}=0.359$
.  We set $\Delta \ln U_{t+1}=0$ at the approximation point, which is close to its post-2007 average anyway.  

\subsection{Results}

\begin{figure}[htbp]
\includegraphics[width=0.6\textwidth]{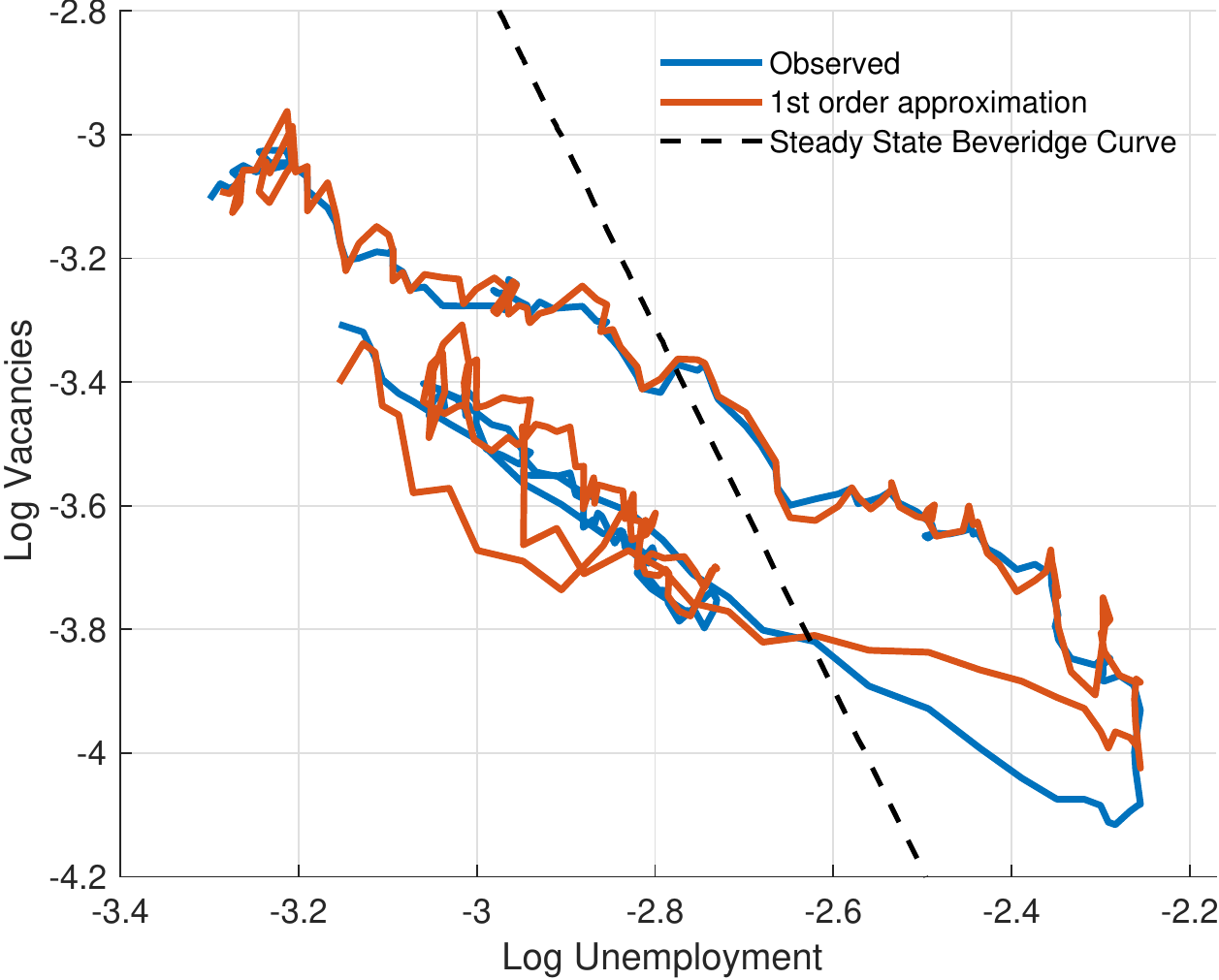}
\caption{Beveridge Curves}
\label{fig:beveridge_and_1st}
\begin{figurenotes}
3 month moving averages of monthly data, 2000-2019.
\end{figurenotes}
\begin{figurenotes}[Source]
CPS, JOLTS.
\end{figurenotes}
\end{figure}

Figure \ref{fig:beveridge_and_1st} plots the (log) observed Beveridge curve, the first order approximation, and the steady-state Beveridge curve.  The approximate Beveridge curve, which includes all the (first order) effects of the shifters, follows the actual curve closely, aside from a brief period near the trough of the Great Recession.  Most importantly, the approximate curve shows nearly the same shift (between recession downswing and recovery) as the observed curve.  The good fit of the linearized curve gives us confidence that our decomposition of the linearized curve will also be accurate for the exact curve.  Appendix \ref{sec:full_decompositions} addresses any lingering concerns about the accuracy of the linearized results by calculating a series of nonlinear decompositions on the exact Beveridge curve.

\begin{figure}
\includegraphics[width=0.7\textwidth]{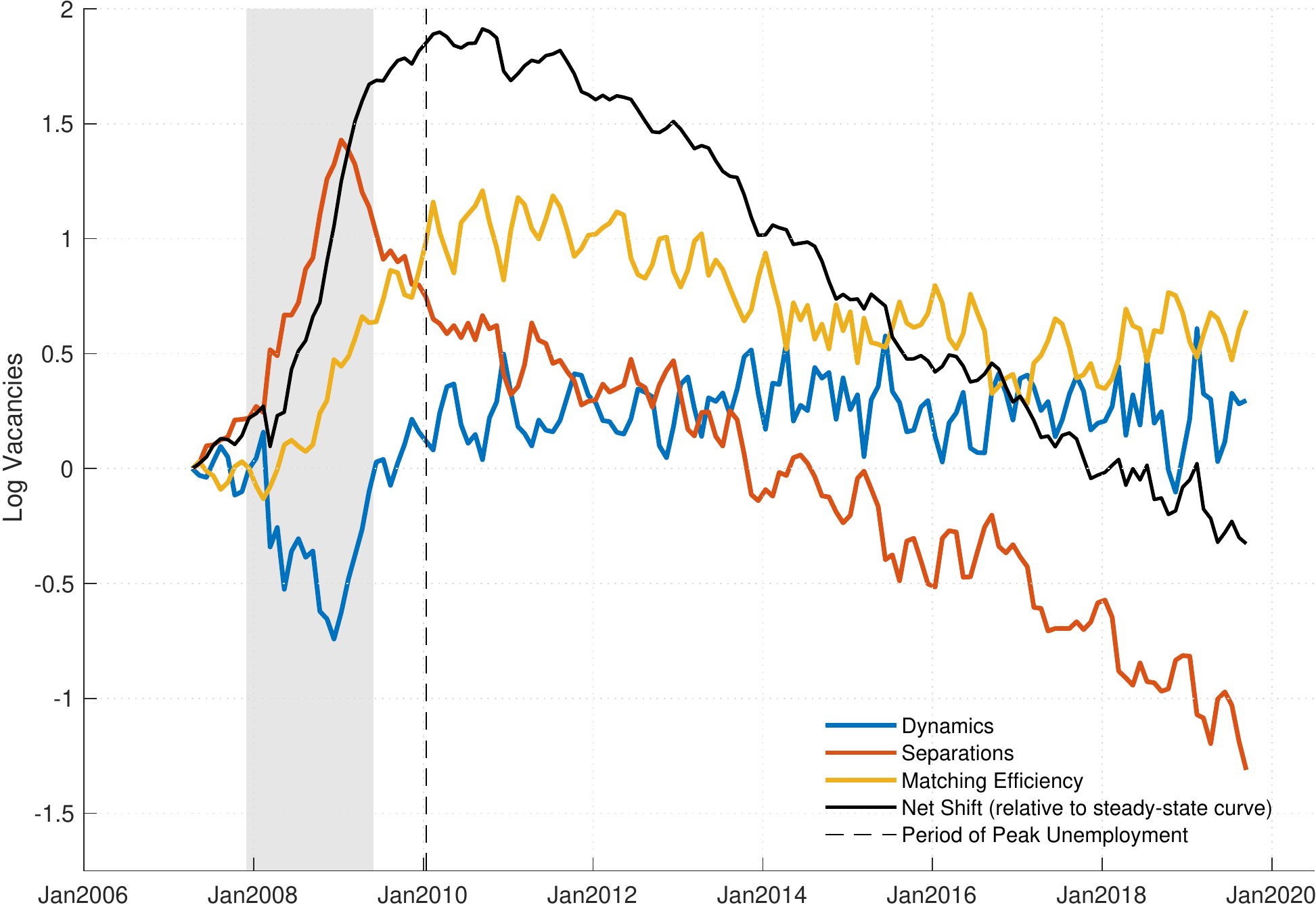}
\caption{Shifters of the Approximate Beveridge Curve}
\label{fig:shifts_time}
\begin{figurenotes}
3 month moving averages of monthly data, 2000-2019. NBER recessions shaded in gray.  Shifters are relative to April 2007 values.
\end{figurenotes}
\begin{figurenotes}[Source]
CPS, JOLTS.
\end{figurenotes}
\end{figure}

Both the actual Beveridge curve and the approximate curve are significantly flatter than the steady state curve.  In log space, the slope of the steady state curve is roughly $-\frac{1-\alpha}{\alpha}=-2.66$, while the slope of the empirical curve is near unity.  The difference in slopes is due to slow variation in the shifters, which pushed vacancies up as the Great Recession took hold, and then pushed vacancies down in the recovery.  Figure \ref{fig:shifts_time} plots the time paths of the three shifter terms in equation \eqref{eq:loglin}, along with the net shift (the black line), all normalized to be zero in April 2007.     The blue line shows the shift in the Beveridge curve attributable to out-of-steady-state dynamics (that is, $ -\frac{\overline{U}}{\alpha\overline{s}(1-\overline{U})}  \Delta \ln U_{t+1}$.)  The red and orange lines similarly show the shifts due to separations and matching efficiency.  

Relative to the pre-Great Recession period (say, 2007), the net effect of the shifters was to move vacancies sharply upward during the recession.  This effect then dissipated very slowly, with the shifters returning to their pre-recession net value only in 2017.  This combined effect explains why the slope of the empirical Beveridge curve is so much flatter than the steady state curve.  We return to this point in Section \ref{sec:slope}.

Turning to each shifter separately, contribution of each factor is complicated and time-varying.  Out-of-steady-state dynamics pushed the Beveridge curve intercept sharply down in the recession, and modestly up in the recovery, more or less the way \cite{pissarides2000} describes.  The contribution of separations is roughly the opposite, raising the intercept sharply, especially late in the recession, and then eventually pushing the intercept down.  Finally, the deterioration in matching efficiency raised the intercept during and after the recession.


Figure \ref{fig:shifts_time} cannot clearly tell us which factors are responsible for the shift in the \emph{empirical} Beveridge curve between the downswing and the upswing of the Great Recession.  To understand that, we need to condition on a level of unemployment and examine the vertical shift evident in Figure \ref{fig:beveridge_and_1st}.

Say that there were two months, $t$ and $t'$, where observed unemployment rates were exactly equal, $U_{t}=U_{t'}$.  Then using equation \eqref{eq:loglin} we could decompose the (approximate) difference in vacancies, $\ln V_{t'}-\ln V_{t}$, as follows:
\begin{align}
\ln V_{t'} & -\ln V_{t}	\approx \nonumber \\ 
	& \underbrace{ -\frac{\overline{U}}{\alpha\overline{s}(1-\overline{U})} \left( \Delta \ln U_{t'+1} - \Delta \ln U_{t+1} \right) }_{\substack{\textit{Shift due to} \\ \textit{Dynamics}}} 
    \underbrace{ +\frac{1}{\alpha(1-\overline{U})}\left(\ln s_{t'}-\ln s_{t} \right) }_{\substack{\textit{Shift due to} \\ \textit{Separations}}}  
    \underbrace{ -\frac{1}{\alpha}\left(\ln \sigma_{t'}-\ln \sigma_{t}\right) }_{\substack{\textit{Shift due to} \\ \textit{Matching Efficiency}}}  \label{eq:loglin_diff}
\end{align}

Equation \eqref{eq:loglin_diff} provides an additive decomposition of the vertical shift in the Beveridge curve.  The portion of $\ln V_{t'}  -\ln V_{t}$ due to, say, differences in matching efficiency between $t$ and $t'$ is just the log difference in matching efficiency, $\ln \sigma_{t'}-\ln \sigma_{t}$, multipled by $1/\alpha$.  The shifts due to dynamics and separations are similar.  The only wrinkle in implementing equation \eqref{eq:loglin_diff} is that we never observe two months with exactly the same unemployment rate, so we linearly interpolate all relevant series.  

\begin{figure}
\includegraphics[width=0.65\textwidth]{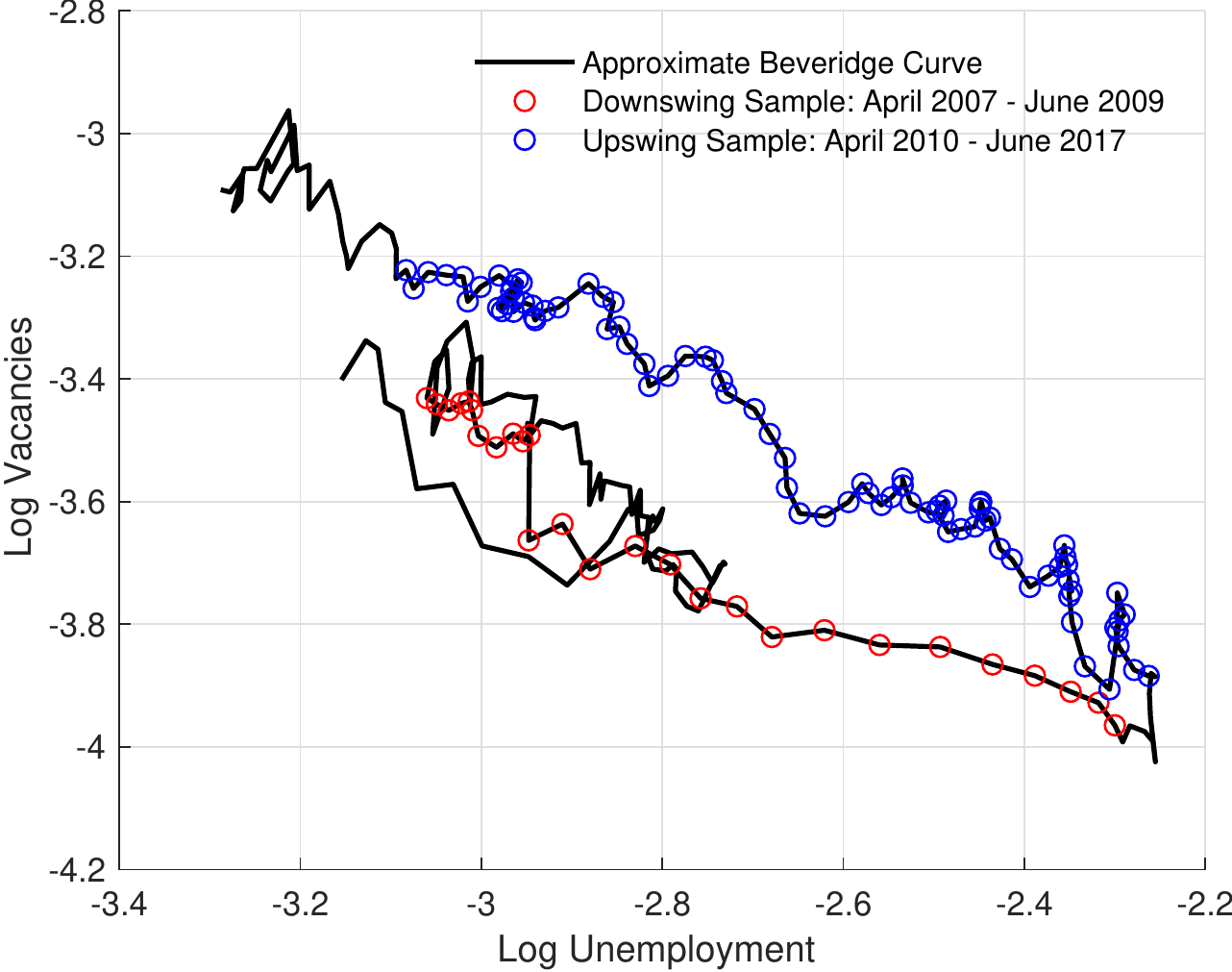}
\caption{Downswing and Upswing Samples}
\label{fig:BC_up_down_samples}
\begin{figurenotes}
3 month moving averages of monthly data, 2000-2019. 
\end{figurenotes}
\begin{figurenotes}[Source]
CPS, JOLTS.
\end{figurenotes}
\end{figure}

As the reference points, we select the unemployment rates observed between April 2007 and June 2009.  These are highlighted in red in Figure \ref{fig:BC_up_down_samples} (the ``downswing sample'').  We compare the downswing sample to the upswing sample, which begins in April 2010 (highlighted in blue).  For each of the downswing points, we calculate the vertical distance between observed vacancies and the (linearly interpolated) upswing vacancy levels.  We also calculate each of the terms in equation \eqref{eq:loglin_diff}.

\begin{figure}
\includegraphics[width=0.65\textwidth]{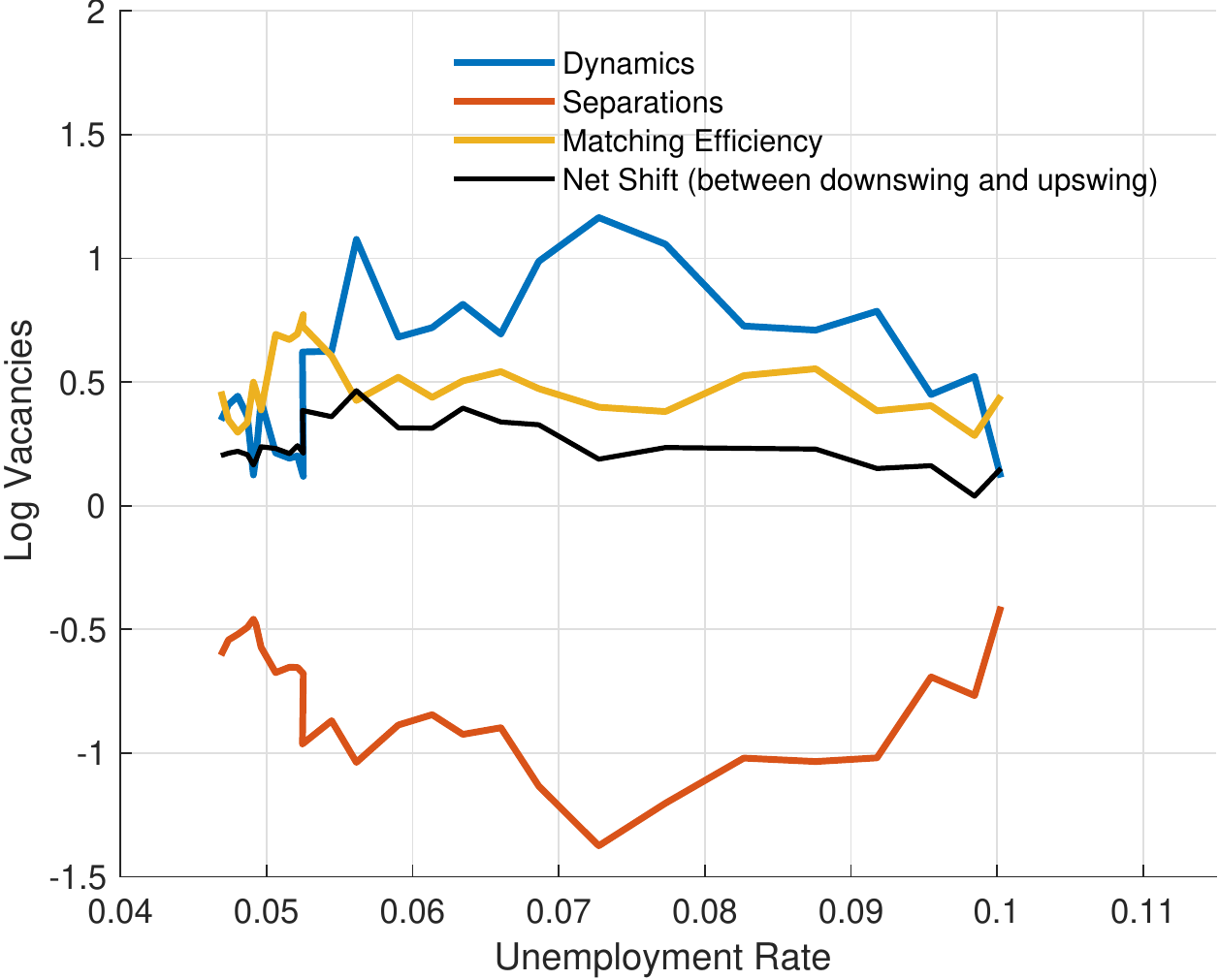}
\caption{Accounting for the Vertical Shift}
\label{fig:shifts_unemp}
\begin{figurenotes}
3 month moving averages of monthly data, 2000-2019.   
\end{figurenotes}
\begin{figurenotes}[Source]
CPS, JOLTS.
\end{figurenotes}
\end{figure}

The result is Figure \ref{fig:shifts_unemp}.  The x-axis is the unemployment rate.  For each unemployment rate, the black line shows the vertical distance between the upswing and downswing samples, as measured in log vacancies.  This is the shift in the Beveridge curve we are trying to explain.  The black line is the sum of the other three lines, which are the contributions in equation \eqref{eq:loglin_diff}.  There are several striking results.  First, the job-separation probability is responsible for a large shift \emph{down} in the Beveridge curve.  This is because separations rose early in the recession, pushing up vacancies, and later fell, making upswing vacancies lower.  This shift is offset by the combined effects of dynamics and matching efficiency, which both pushed the curve up on net.

Interestingly, out-of-steady-state dynamics played a prominent role, with a contribution larger than that of matching efficiency over much of the range.  This specific result is consistent with \cite{CET2015}'s argument that, because the Great Recession was so large and so sudden, dynamics can produce a realistic loop in the Beveridge curve.  However, their analysis ignores the separation probability and matching efficiency, which are at least as important for understanding what happened.  In particular, matching efficiency more than accounts for the net shift across most of the range, so without a change in matching efficiency the Beveridge curve would have shifted \emph{down}, not up.  

To summarize, all three of the factors we consider shifted the Beveridge curve in non-trivial ways.  The vertical shift in the empirical Beveridge curve is the net result of out-of-steady state dynamics and matching efficiency both shifting the curve up, an effect which is partially offset by a large negative contribution from the separation probability.  The time paths of these shifters are complicated and non-monotonic, leading the slope of the empirical Beveridge curve to differ from the model-implied steady-state curve.  We now turn to this result in more detail.

\subsection{The Slope of the Beveridge Curve}\label{sec:slope}

Recent innovative work by \cite{michaillat2019} (MS) has emphasized the importance of the Beveridge curve slope for welfare and the natural rate of unemployment.  In this section we show how our measurement methods relate to their results.

In many models with a matching function (e.g., \cite{shimer2005}), the Beveridge curve describes the possible steady-state values of vacancies and unemployment.  In short, an economy that sustains a low level of unemployment \emph{must} have more vacancies in equilibrium, and vice versa.  MS point out that this relationship can be used to estimate the welfare-maximizing level of unemployment in a particularly simple and general way.  They note that a social planner will seek to equalize the costs of additional vacancies to the costs of additional unemployment.  In other words, the social planner will seek the location on the Beveridge curve where the marginal cost of additional unemployment equals the social value of the resulting reduction in vacancies.  This point then defines the natural rate of unemployment, and the difference between observed unemployment and natural rate is the unemployment gap.  MS use estimates of the costs of vacancies, the costs of unemployment, and the slope of the Beveridge curve to make their calculations.

MS measure the slope of the Beveridge curve by estimating regressions of $V_{t}$ on $U_{t}$ in periods where the Beveridge curve appeared stable (dropping the troughs of recessions, for example.)  As we show above, these observed slopes reflect both (1) movements along a stable Beveridge curve (changes in $U_{t}$ for fixed separations, matching efficiency and dynamics) and (2) time variation in the shifters. This second factor can distort the empirical Beveridge curve relative to the planner-relevant, steady-state curve.  For example, consider a bare bones model where the separation probability and matching efficiency are exogenous processes, possibly correlated with the aggregate productivity shock.  Such a model fits in our framework (and that of MS), and could produce the observed data, including the empirical Beveridge curve and the paths of the shifts.  However, a planner, facing such an economy, would not look to the empirical Beveridge curve to estimate the unemployment-vacancy tradeoff.  The reason is that the empirical curve include the effects of the (purely cyclical) shifters, while the planner is interested in long-run, steady state relationships.  The correct slope for the planner comes from the linearized curve \eqref{eq:loglin}, which treats the shifters as fixed:

\begin{align}\label{eq:bc_slope}
    -\left(\frac{\overline{U}}{\alpha\left(1-\overline{U}\right)} +\frac{1-\alpha}{\alpha}\right) &\approx -\frac{1-\alpha}{\alpha}
\end{align}

\noindent and is determined by the shape of the matching function. The planner would make decisions based on the steady-state curve in Figure \ref{fig:beveridge_and_1st}, not the empirical curve.  Thus, in this toy example the empirical Beveridge curve does not directly give us the planner-relevant, long-run relationship we seek.  

The key question is whether the planner should incorporate the effects of the shifters when making a choice about the long-run level of unemployment. Clearly, out-of-steady-state dynamics are fundamentally transitory, so the planner should always purge the Beveridge curve of their effect.   However, it is possible that the separation probability and matching efficiency are, to some extent, functions of the long-run level of unemployment (unlike in the toy example above).  In this case the planner should not remove (all of) their influence when calculating the vacancy-unemployment tradeoff.  

Determining the exact nature of the variation in separations and matching efficiency is well beyond the scope of this paper.  Instead, we provide an example to demonstrate that these issues can have an economically meaningful impact on welfare calculations.  From equation \eqref{eq:bc_slope}, the slope of the steady-state curve (treating the shifters as fixed) is very close to $-\frac{1-\alpha}{\alpha}$.  Averaging together the two estimates of $\alpha$ in Table \ref{tab:match_fun_table}, we set set $\alpha=0.3$, implying a Beveridge curve slope of $-2.33$.   This is far steeper than the estimates of MS, which are around $-0.9$ for the same period.  

\begin{figure}
\includegraphics[width=0.6\textwidth]{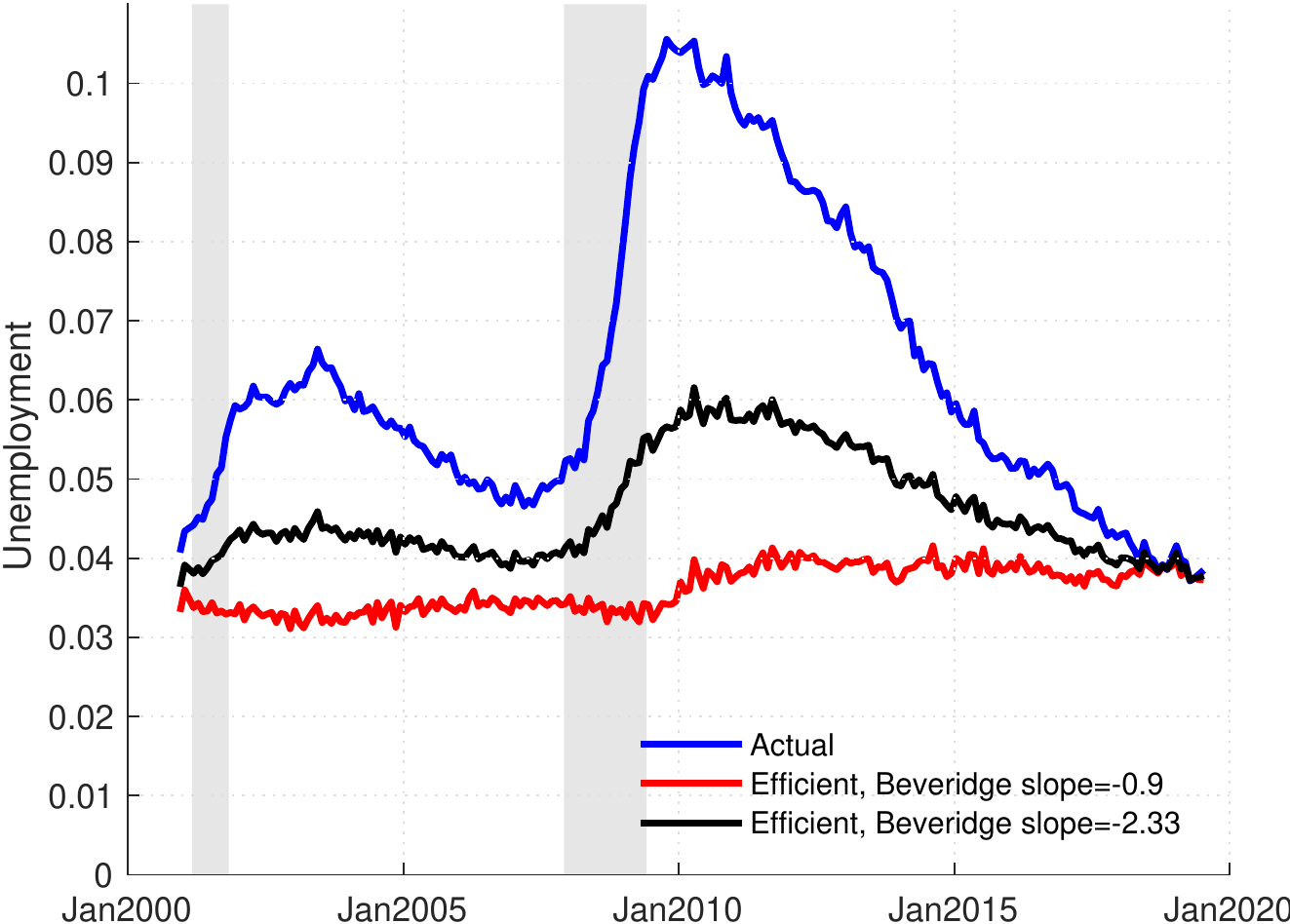}
\caption{Efficient Unemployment Based on the Beveridge Tradeoff}
\label{fig:MS}
\begin{figurenotes}
Calculations based on 3 month moving averages of monthly data, 2000-2019. 
\end{figurenotes}
\begin{figurenotes}[Source]
CPS, JOLTS.
\end{figurenotes}
\end{figure}

We can calculate the efficient levels of unemployment using equation (5) from MS, based on our two estimates of the Beveridge slope ($-2.33$ and $-0.9$).  In both cases we use MS's preferred values for the costs of vacancies and unemployment.  Figure \ref{fig:MS} shows the results (this figure is comparable to Figure 3 Panel D in \cite{michaillat2019}.)  The blue line is the actual unemployment rate.  The red line shows the efficient level of unemployment according to MS's calibration, with a Beveridge slope of $-0.9$.  The black line shows the efficient level of unemployment using our preferred Beveridge slope of $-2.33$.  It is evident that the steeper Beveridge curve significantly raises the efficient level of unemployment, as reducing unemployment with a steep Beveridge curve is more costly in terms of vacancies.  In our calibration the natural rate of unemployment fluctuates between about 4 percent and 6 percent, near the range of other estimates including the Congressional Budget Office (CBO)'s short-term natural rate of unemployment. Notably, our calibrated estimate moves very similarly to the CBO's estimate during the post Great-recession period.

Our results suggest that careful work is needed to disentangle which features of the Beveridge curve the planner should care about.  These choices have real consequences for the measurement of efficiency, as Figure \ref{fig:MS} shows.  One approach is to specify a more complete model, which explicitly links separations and matching efficiency to the rest of the economy.  With such a model in hand, one could determine the planner-relevant Beveridge curve slope.

\section{Historical Recessions}\label{sec:historical_recessions}

\begin{figure}
  \centering
  \subfigure[]{\includegraphics[width=0.42\textwidth]{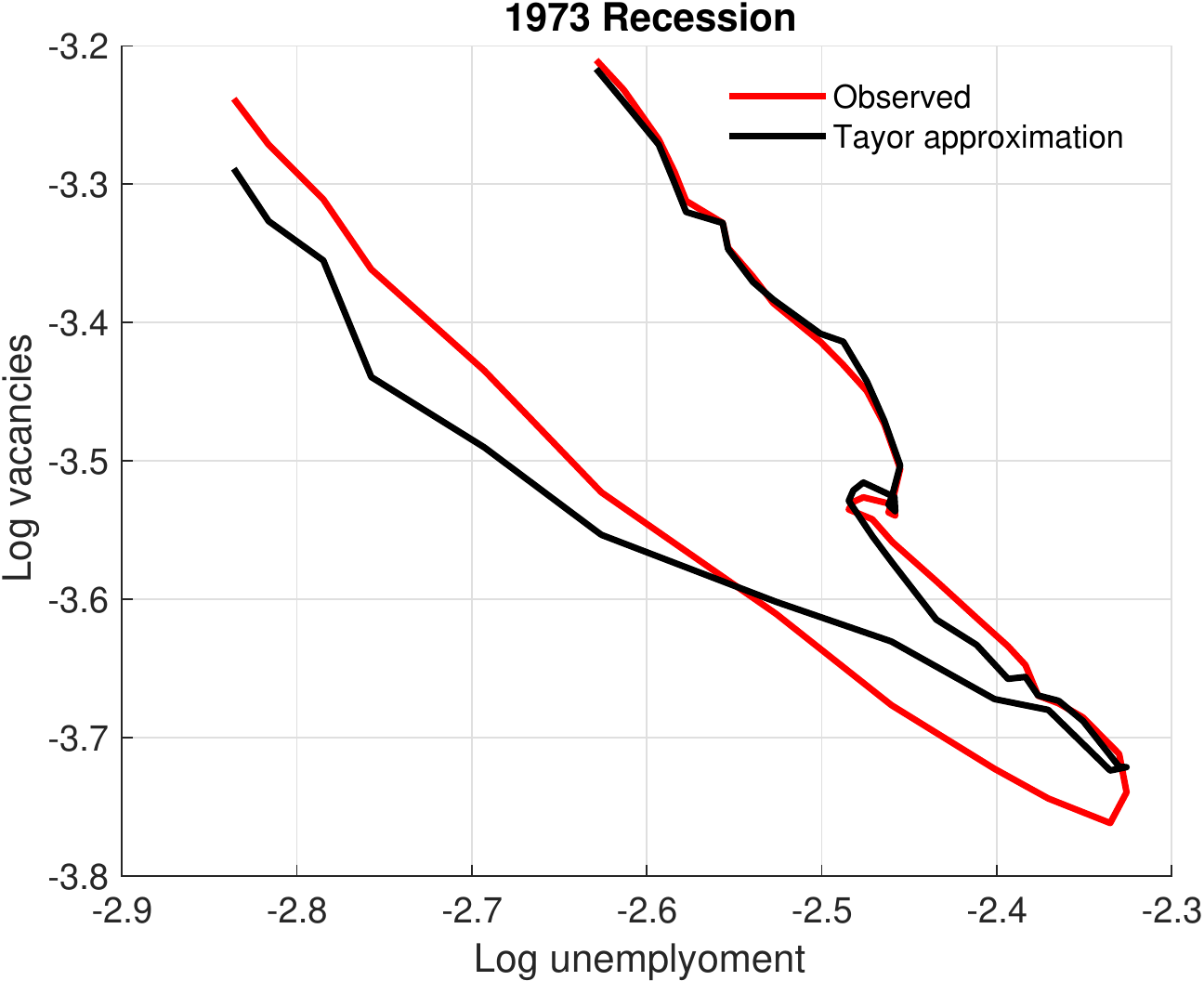}}
  \subfigure[]{\includegraphics[width=0.42\textwidth]{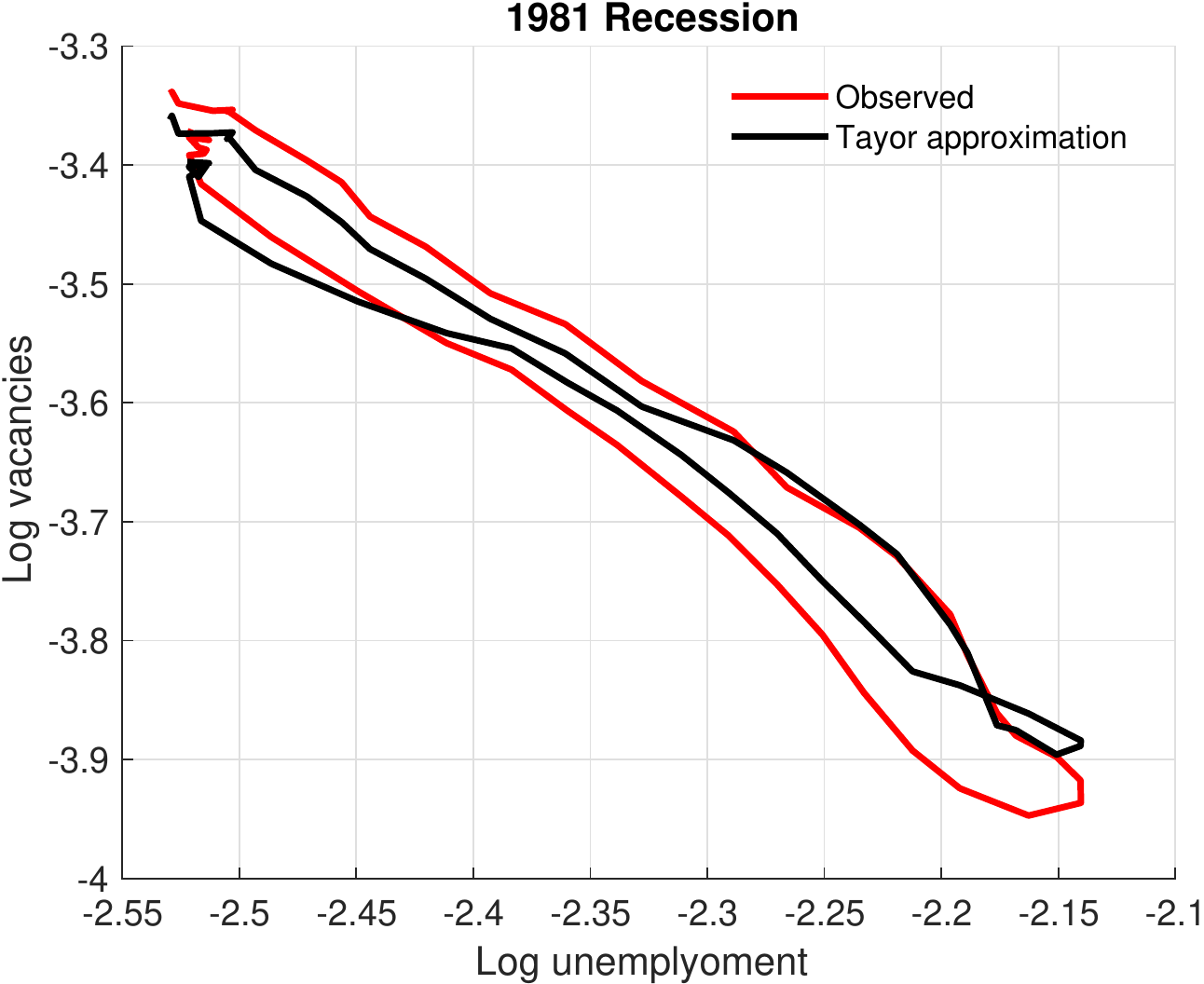}}  \subfigure[]{\includegraphics[width=0.45\textwidth]{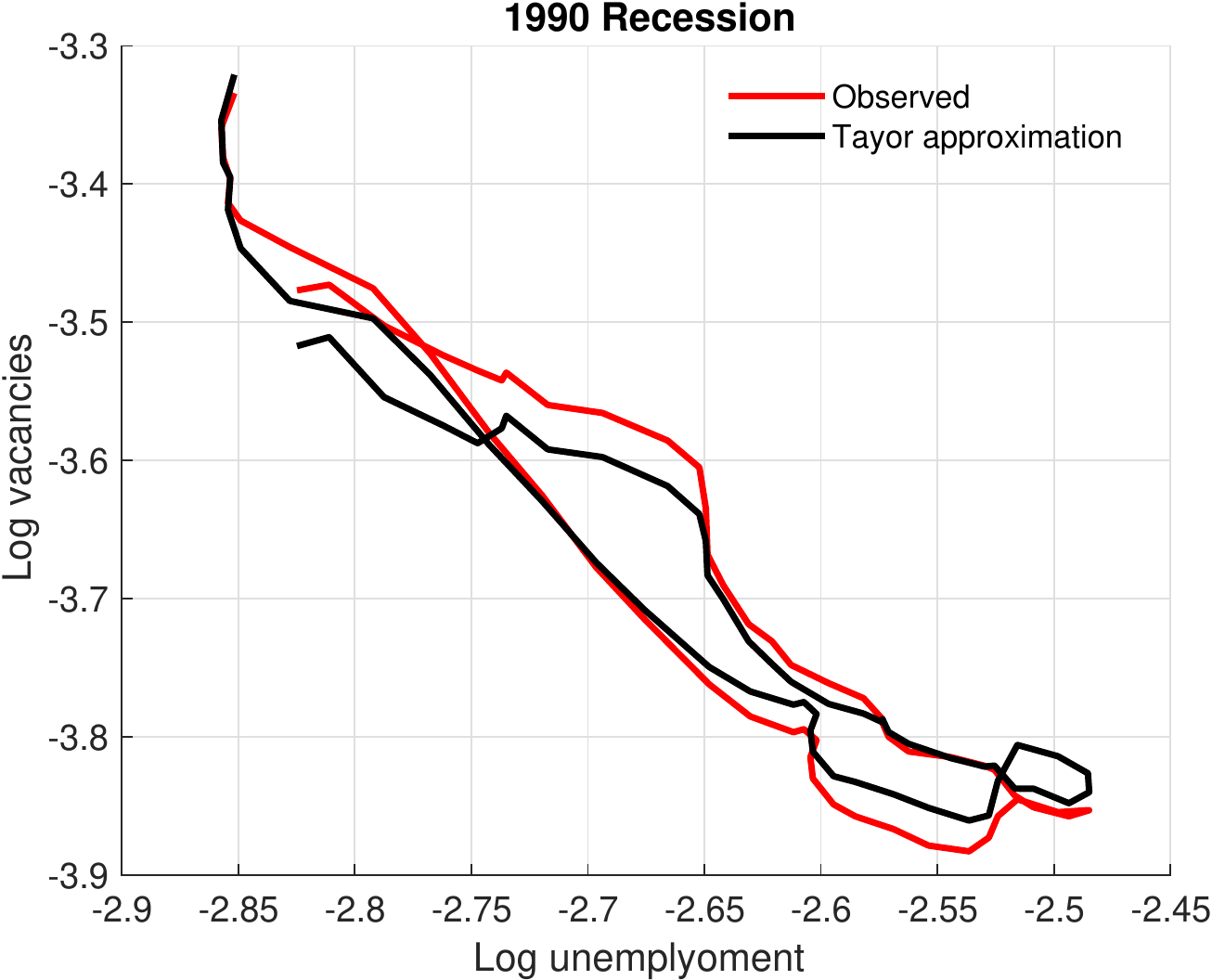}}
  \subfigure[]{\includegraphics[width=0.42\textwidth]{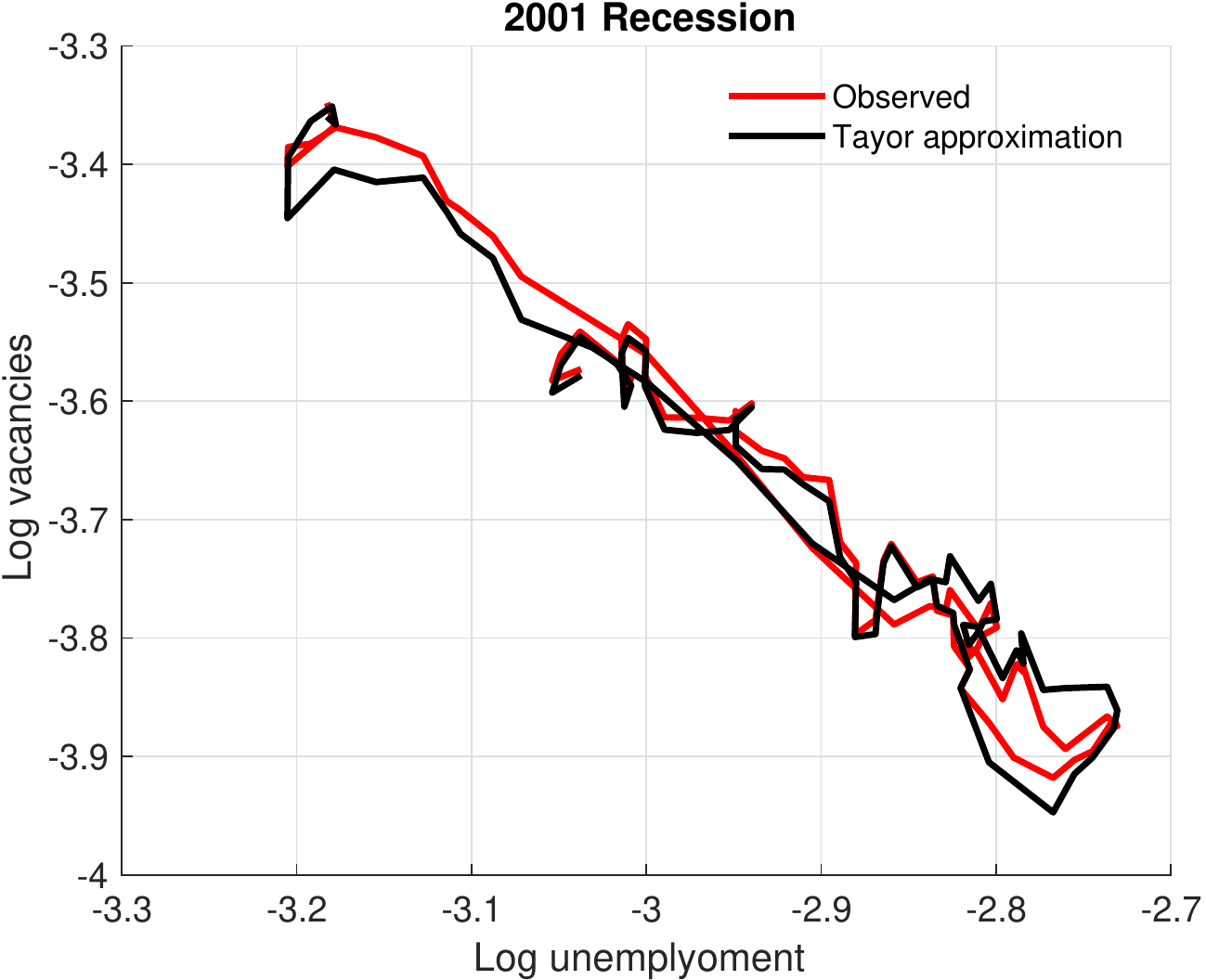}}
  \caption{Observed and Approximate Beveridge Curves}
  \label{fig:hist_taylor}
\begin{figurenotes}
Observed and approximated Beveridge curves for historical downturns.  Log-linear Taylor approximations are taken about averages for the periods plotted.  
\end{figurenotes}
\begin{figurenotes}[Source]
CPS, \cite{barnichon2010}, and authors' calculations.
\end{figurenotes}
\end{figure}

We can also use our framework to analyze recessions prior to the Great Recession.  In terms of data, the only change is that up through 2016 we use the composite vacancy series from \cite{barnichon2010} instead of JOLTS.  After 2016 we continue the series by splicing on the JOLTS series.  For four historical labor market downturns, we calculate the log-linearized Beveridge curve, as in Section 4.  For each episode the curve is linearized around the local mean, to ensure a good fit.  Figure \ref{fig:hist_taylor} compared the observed and linearized Beveridge curves.  The fit is generally good, although some of the linearized Beveridge curves show less of a shift, or counter-clockwise loop, than their observed counterparts.  We view this as a topic for further investigation

\begin{figure}
  \centering
  \subfigure[]{\includegraphics[width=0.45\textwidth]{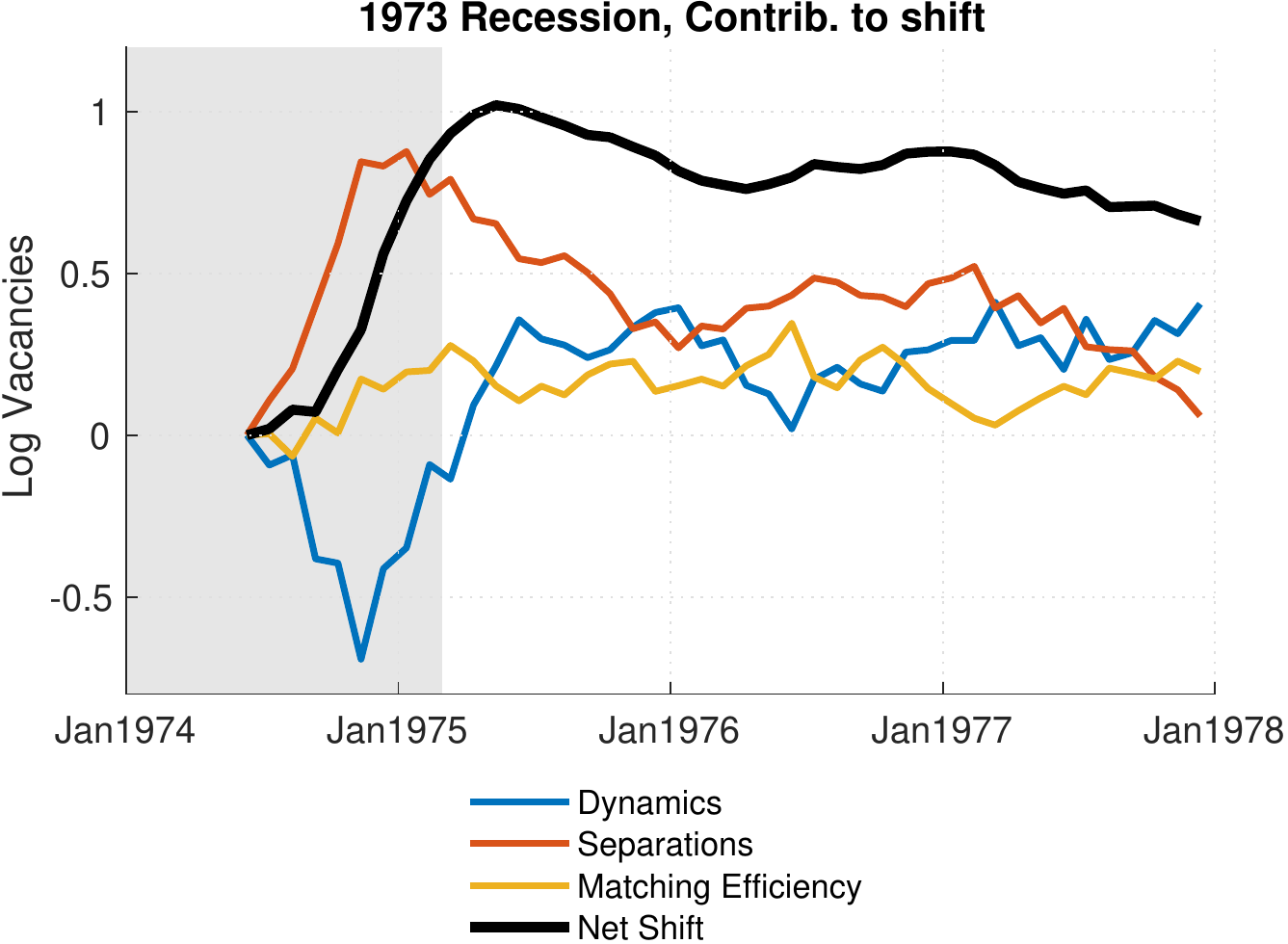}}
  \subfigure[]{\includegraphics[width=0.45\textwidth]{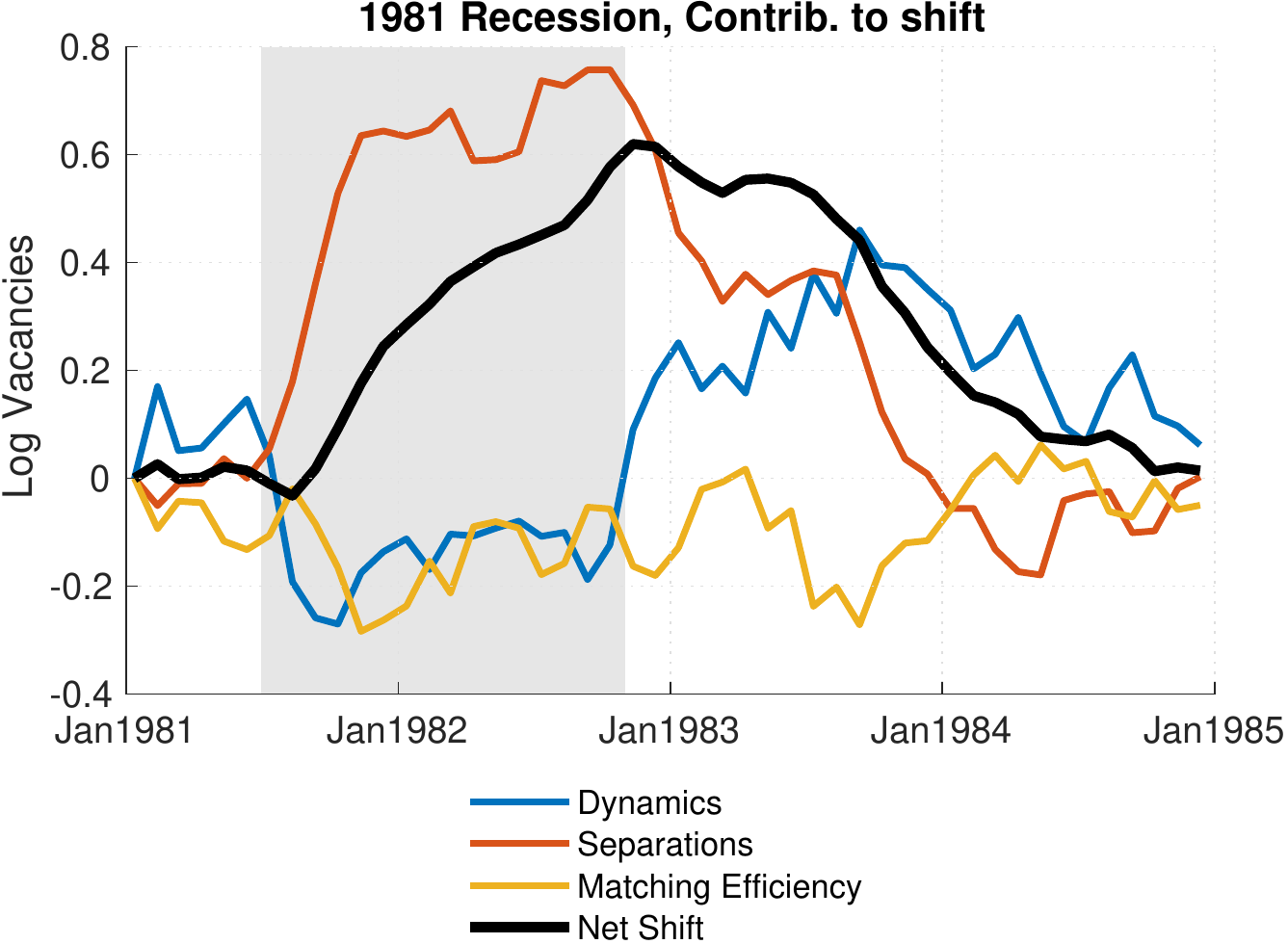}}  \subfigure[]{\includegraphics[width=0.45\textwidth]{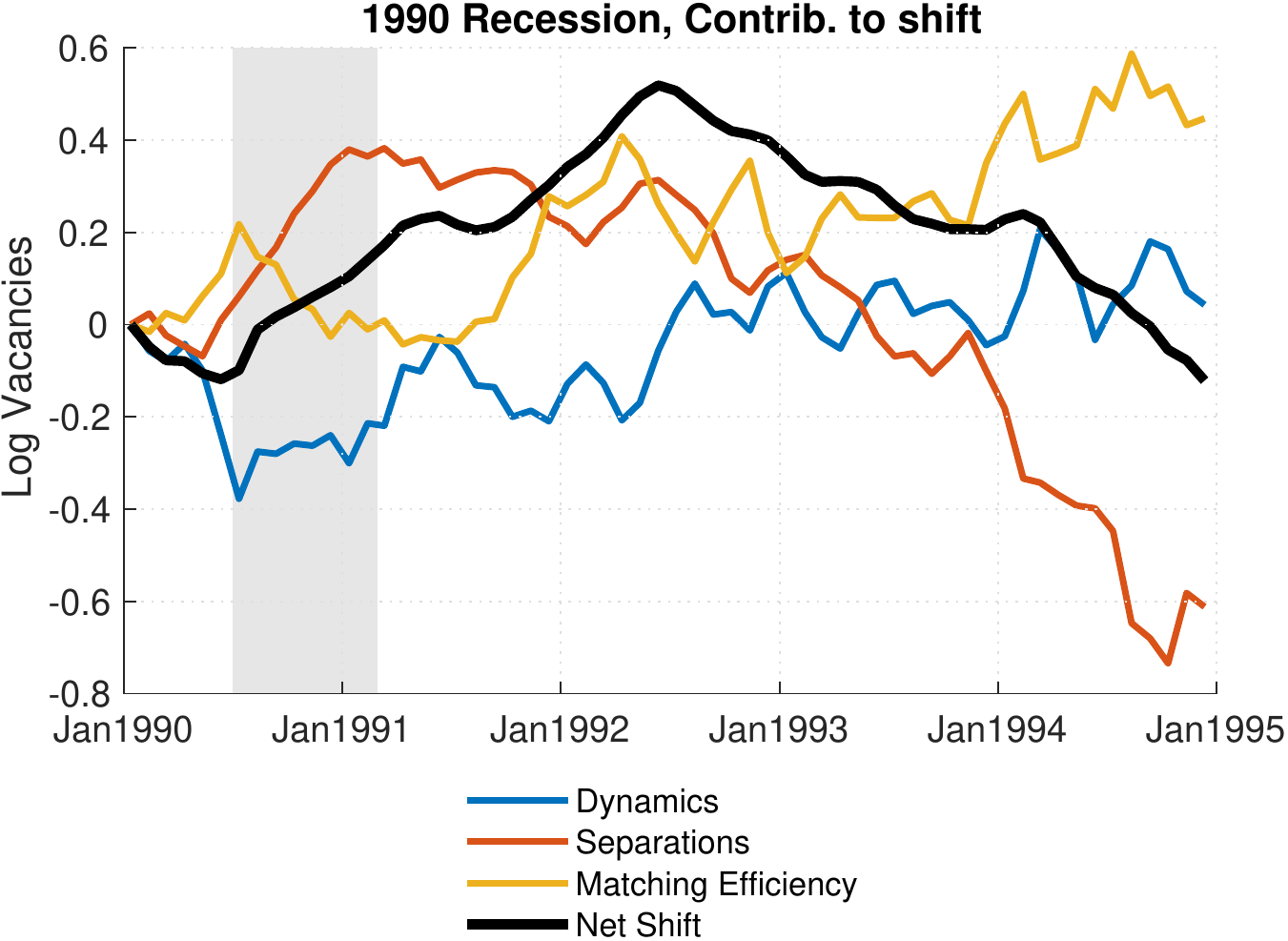}}
  \subfigure[]{\includegraphics[width=0.45\textwidth]{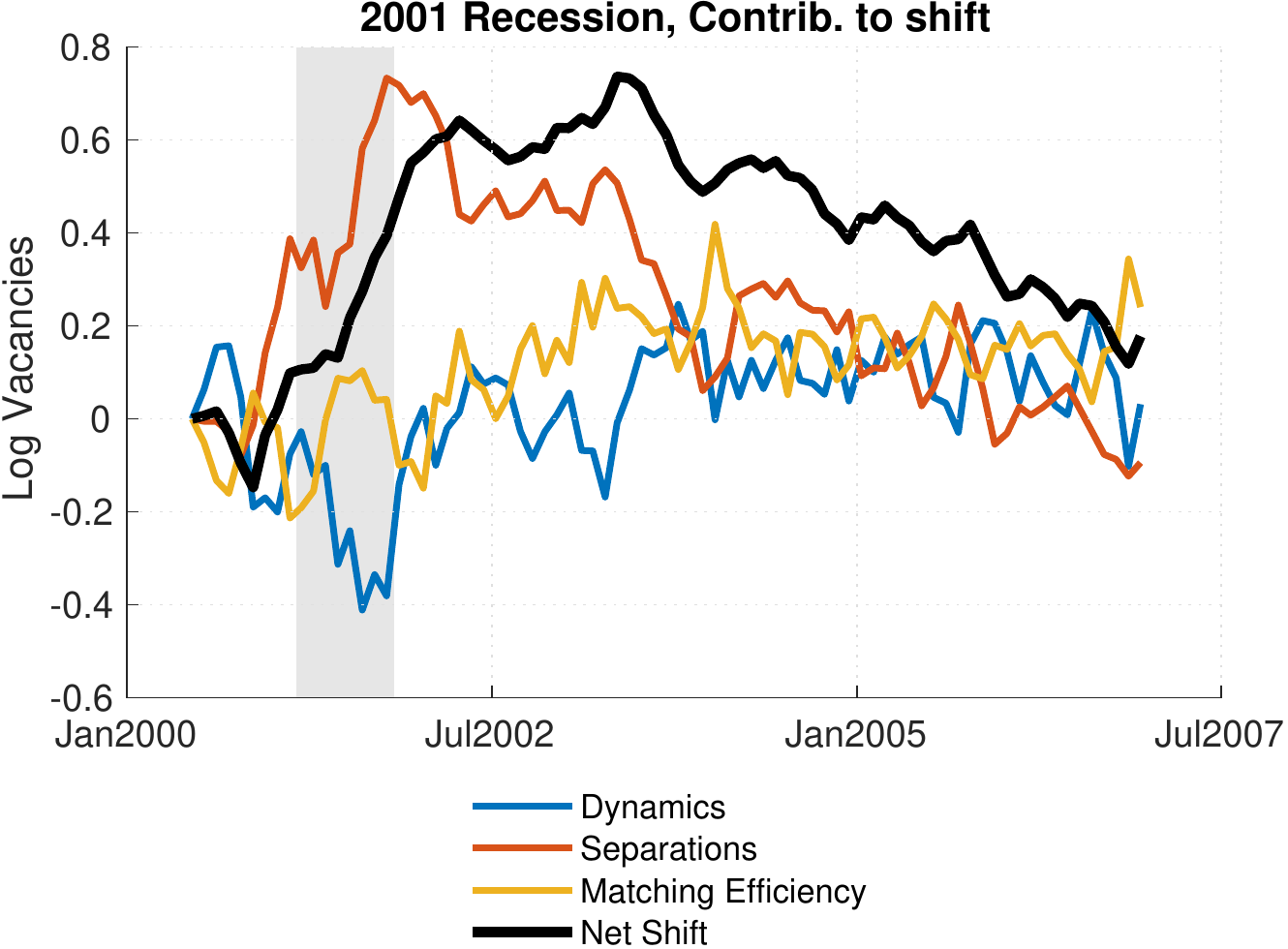}}
  \caption{Decompositions of Approximate Beveridge Curves}
  \label{fig:hist_decomp_time}
\begin{figurenotes}
Solid black line shows the time path of the net shift of the (log-linearized) Beveridge curve intercept.  NBER recessions shaded in gray.   
\end{figurenotes}
\begin{figurenotes}[Source]
CPS, \cite{barnichon2010}, and authors' calculations.
\end{figurenotes}
\end{figure}

With the linearized Beveridge curves in hand, we can read off the implied contribution of each factor to the shift in the curve at every point in time.  Figure \ref{fig:hist_decomp_time} presents the historical versions of Figure \ref{fig:shifts_time}: the net shift in the Beveridge curve, and the contributions, as functions of time.  It is apparent that in each recession the Beveridge curve intercept began shifting up at the onset of the recession, and slowly drifted down once unemployment began falling.  Rising separations usually drove this upward shift, partially offset by out-of-steady-state dynamics.

It can be seen that in all recessions, out-of-steady-state dynamics shifted the Beveridge curve significantly down in the initial stages (the light blue line is below zero) and generally up in the recovery (the bold blue line is above zero).  Interestingly, this shift is partially offset by the contribution of separations, which (as in the Great Recession) tend to push Beveridge curve sharply upward in  the initial stages of a recession and more moderately upward afterward.  Thus the changes in the job-separation probability tend to flatten the observed Beveridge curve, and cancel out some of the counter-clockwise loop that out-of-steady-state dynamics induce.  

In most previous recessions, changes in matching efficiency had little impact, and were swamped by changes in the other factors.  The 1990 recession appears to be an exception here. During the 1990 recession and the recovery period, the deterioration in matching efficiency continued to push the Beveridge curve up, which is quite similar to what happened in the Great Recession. In fact, the two recessions are similar to each other in a sense that long-term unemployment continued to increase substantially after the recession was over. This suggests  that mismatch or related factors might have been an important driver in the rise of long-term unemployment in the two recession episodes. We view this line of reasoning as a topic for future research.

The tentative conclusion is that the Great Recession was exceptional, insofar as the drop in matching efficiency had first-order effects on the Beveridge curve (with the possible exception of the early 1990s recession).  In previous recessions matching efficiency usually played little role.  However, the modest counter-clockwise loops in previous recession were not simply the product of modest out-of-steady-state dynamics, but were the net result dramatic dynamics being offset by large contributions from the separations margin.  Out-of-steady-state dynamics and the separations margin played critical roles in all the recessions examined here.

\section{Three State Model} \label{sec:3state}

The results so far have assumed that all workers are either employed or unemployed.  This is a significant simplification, since empirically flows into and out of the labor force are important for understanding total hires and evolution of unemployment.  In this section we add a participation margin and discuss the robustness of our results in the expanded model.

\subsection{Model}

The population is still normalized to unity, but we add a nonemployment state.  Let $N_{t}$ is the stock of nonemployed, so that $E_{t}+U_{t}+N_{t}=1$.   Consider the law of motion for unemployment when workers can move into and out of the labor force: 
\begin{align} \label{eq:3state_lom}
    \Delta U_{t+1} &= E_{t}eu_{t} + N_{t}nu_{t}-U_{t}un_{t}-U_{t}ue_{t}
\end{align}
The transition rate from nonemployment to unemployment in month $t$ is $nu_{t}$.  The terms $un_{t}$ and $ue_{t}$ are similarly defined, with $eu_{t}$ replacing $s_{t}$ for consistency.  The law of motion for nonemployment is symmetric:
\begin{align} \label{eq:3_N_state_lom}
    \Delta N_{t+1}&=E_{t}en_{t}+U_{t}un_{t}-N_{t}ne_{t}-U_{t}un_{t}
\end{align}
Summing equations \eqref{eq:3state_lom} and \eqref{eq:3_N_state_lom} yields an expression involving total hires ($H_t=N_{t}ne_{t}+U_{t}ue_{t}$)
\begin{align} \label{eq:3_summed_lom}
    \Delta U_{t+1}+ \Delta N_{t+1}&=E_{t}eu_{t} + E_{t}en_{t}+H_{t}
\end{align}
where the flows between unemployment and nonemployment have canceled.

We can write the matching function as 
\begin{align} \label{eq:3state_matching}
    H_{t}&=\sigma_{t}(U_{t}+\xi_{t}^{N}N_{t})^{1-\alpha}V_{t}^{\alpha}
\end{align}
where $\xi_{t}^{N}$ is the search effort of the nonemployed relative to the unemployed.  Thus the effective mass of searchers is $U_{t}+\xi_{t}^{n}N_{t}$ and $\sigma_{t}$ continues to represent reduced-form matching efficiency.

Combining equations \eqref{eq:3_summed_lom} and \eqref{eq:3state_matching}, and assuming balanced matching (that is, hires from unemployment are a share $\frac{U_{t}}{U_{t}+\xi_{t}^{N}N_{t}}$ of total hires), we have the following expression for vacancies: 
\begin{equation} \label{eq:3state_V}
    V_{t}=\left[\frac{(1-U_{t}-N_{t})(eu_{t}+en_{t})-\Delta U_{t+1}-\Delta N_{t+1}}{\sigma_{t}(U_{t}+\xi^{N}_{t}N_{t})^{1-\alpha}}\right]^{1/\alpha}
\end{equation}
When the non-employed can participate in job search, it is more sensible to think of a Beveridge curve which relates vacancies to \emph{searchers} (both unemployed and nonemployed) instead of unemployment.  To this end, we make two substitutions.  First, we define the pool of searchers $S_{t}$ as 
\begin{equation} \label{eq:3state_S}
    S_{t}=U_{t}+\xi^{N}_{t}N_{t}.
\end{equation}
Second, we define the pool of ``truly nonemployed'' as 
\begin{equation} \label{eq:3state_Ntilde}
    \tilde{N}_{t}=\left(1-\xi^{N}_{t}\right)N_{t}.
\end{equation}
While we take no stand on whether $\xi^{N}_{t}$ is the fraction of nonemployed who search or the search effort of each nonemployed relative to the unemployed, the former interpretation is convenient here.   Note that if $\xi^{N}_{t}=1$ all the nonemployed search and  $\tilde{N}_{t}=0$.  Using $S_{t}$ and $\tilde{N}_{t}$, we can write \eqref{eq:3state_V} as 
\begin{equation} \label{eq:3state_Vnew}
    V_{t}=\left[\frac{\left(1-S_{t}-\tilde{N}_{t}\right)x_{t}-\Delta S_{t+1}-\Delta\tilde{N}_{t+1}}{\sigma_{t}S_{t}^{1-\alpha}}\right]^{\frac{1}{\alpha}}
\end{equation}
where $x_{t}=eu_{t}+en_{t}$ is the total job-separation probability.  Log-linearizing yields

\begin{align} \label{eq:3state_Vlinear}
    \ln V_{t}&=-\frac{1}{\alpha}\left[\ln\sigma_{t}-\ln\sigma_{0}\right]  \nonumber \\
            &-\left\{ \frac{\left(1-\alpha\right)}{\alpha}+\frac{1}{\alpha}\frac{S_{0}}{\left(1-S_{0}-\tilde{N}_{0}\right)}\right\} \left[\ln S_{t}-\ln S_{0}\right] \nonumber \\
            &-\left\{ \frac{1}{\alpha}\frac{S_{0}}{\left(1-S_{0}-\tilde{N}_{0}\right)x_{0}}\right\} \left[\Delta\ln S_{t+1}\right] \nonumber \\
            &-\left\{ \frac{1}{\alpha}\frac{\tilde{N}_{0}}{\left(1-S_{0}-\tilde{N}_{0}\right)}\right\} \left[\ln\tilde{N}_{t}-\ln\tilde{N}_{0}\right] \nonumber \\
            &-\left\{ \frac{1}{\alpha}\frac{\tilde{N}_{0}}{\left(1-S_{0}-\tilde{N}_{0}\right)x_{0}}\right\} \left[\Delta\ln\tilde{N}_{t+1}\right] \nonumber \\
            &+\frac{1}{\alpha}\left[\ln x_{t}-\ln x_{0}\right]
\end{align}

Like equation \eqref{eq:CET_identity}, equation \eqref{eq:3state_Vnew} can be used to analyze the Beveridge curve.  This decomposition, naturally, has more shifters than the two-state model.  In this model movements along the Beveridge curve are captured by the $\ln S_{t}-\ln S_{0}$ term, since the curve is defined in terms of searchers, not merely the unemployed.  The effects of matching efficiency and separations still appear, on the first and last lines of equation  \eqref{eq:3state_Vnew} respectively.  Finally, there are now two out-of-steady state terms, $\Delta\ln S_{t+1}$ and $\Delta\ln\tilde{N}_{t+1}$, as well as a term capturing the level of non-searchers, $\ln\tilde{N}_{t}-\ln\tilde{N}_{0}$.     Not all of these terms have a transparent interpretation, but as we shall see below, many of them are not quantitatively important either. 


\subsection{Data}

To implement the three state model, we need data on the terms appearing in equation \eqref{eq:3state_V}.  We obtain the stocks of employed, unemployed, and nonemployed from the CPS labor force status flows.\footnote{Accessible at \url{https://www.bls.gov/webapps/legacy/cpsflowstab.htm}.}  We normalize these stocks to satisfy $E_{t}+U_{t}+N_{t}=1$ in all periods.  The transition rates $eu_{t},nu_{t},un_{t},ue_{t}$ are also taken from the labor force status flows.  These transition rates are not exactly consistent with the stocks, due to missing month-to-month linkages and sample rotation.  We iteratively rake the rates until they are consistent with the stocks.  This results in very small adjustments to the transition rates.

Under the assumption of balanced matching, $\xi_{t}^{N}$ can be identified by the ratio of transition rates to employment: 
 \begin{align*}
    \xi_{t}^{N} &= \frac{ne_{t}}{ue_{t}}
\end{align*}

Finally, $\alpha$ and $\sigma_{t}$ can be identified by the matching function regression, using $U_{t}+\xi_{t}^{N}N_{t}$ as the population of effective searchers.  

\subsection{Results}

\begin{figure}
    \includegraphics[width=0.60\textwidth]{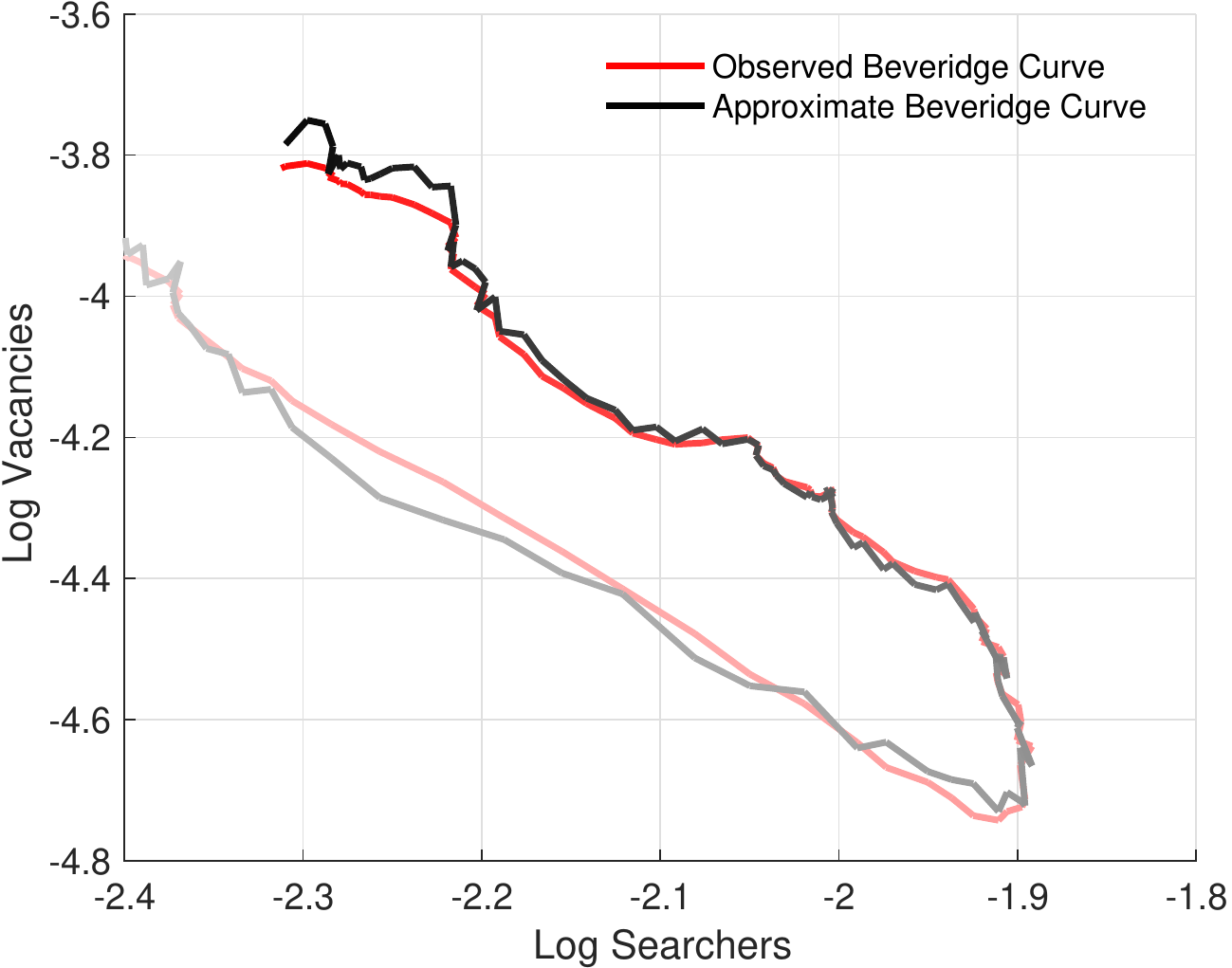}
    \caption{Three State Approximate Beveridge Curve}
    \label{fig:3s_beveridge_and_1st}
    \begin{figurenotes}
    4 month moving averages of monthly data, 2000-2019.  More recent months are shaded darker.  
    \end{figurenotes}
    \begin{figurenotes}[Source]
    CPS, JOLTS.
    \end{figurenotes}
\end{figure}

Figure \ref{fig:3s_beveridge_and_1st} shows that, as with the two state model, the three state approximate Beveridge curve is a good approximation of the observed curve.  Here ``searchers'' are the pool of actively searching workers, $U_{t}+\xi^{N}_{t}N_{t}$.  To show the direction of time, more recent periods are shaded darker.

\begin{figure}
\includegraphics[width=0.6\textwidth]{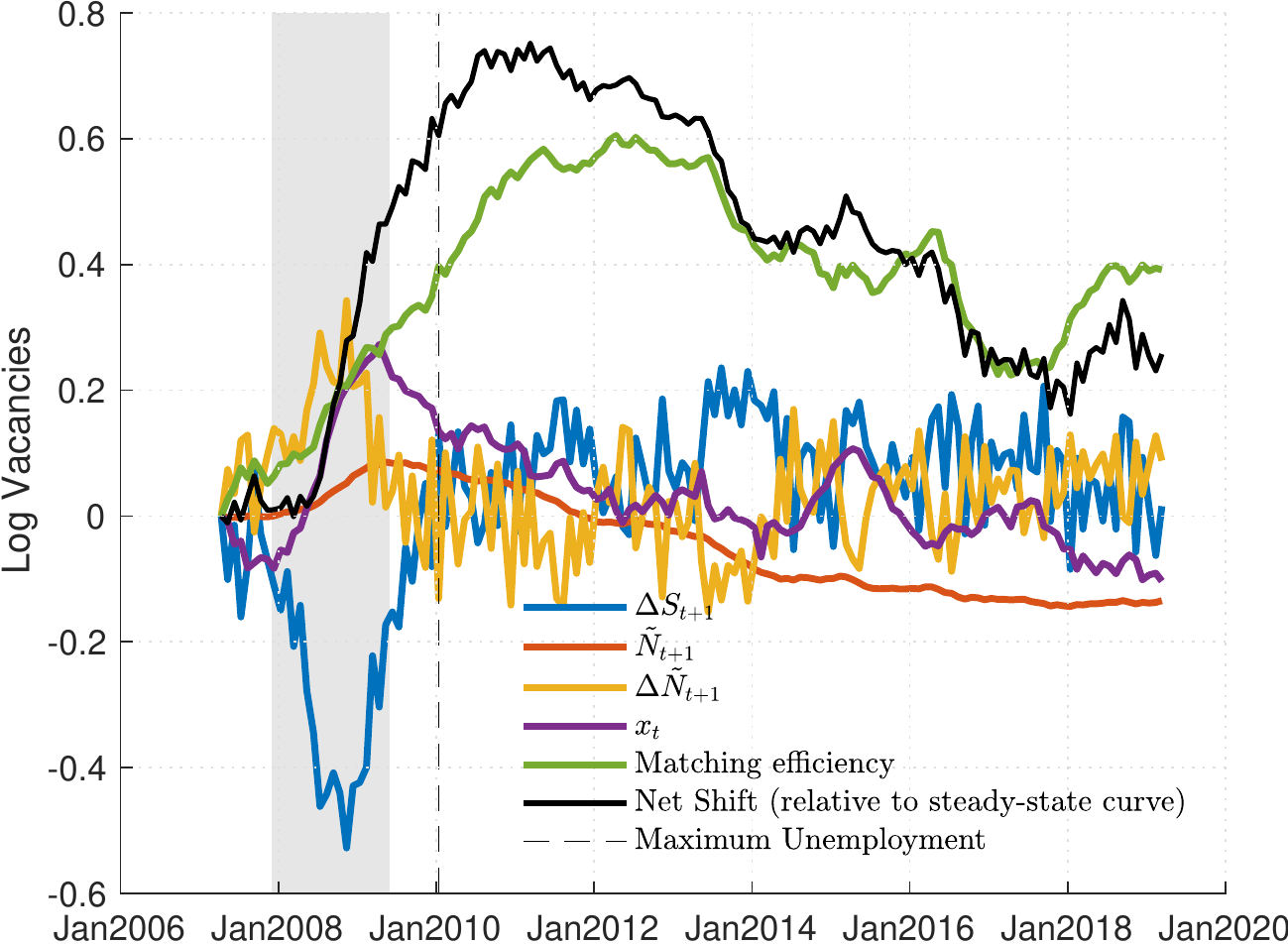}
\caption{Three State Model - Shifters of the Approximate Beveridge Curve}
\label{fig:shifts_time_3s}
\begin{figurenotes}
4 month moving averages of monthly data, 2007-2019. NBER recessions shaded in gray.  Shifters are relative to April 2007 values.
\end{figurenotes}
\begin{figurenotes}[Source]
CPS, JOLTS.
\end{figurenotes}
\end{figure}

Figure \ref{fig:shifts_time_3s} shows the shifters as a function of time, similar to Figure  \ref{fig:shifts_time} the story is similar to the two-state model.  Matching efficiency slowly and steadily pushed the Beveridge curve upwards during and after the Great Recession. The separation probability, $x_{t}$, pushed the Beveridge curve up during the recession, but this was short-lived.  The out-of-steady-state dynamics terms, on net, pushed the curve down, though interestingly the $\Delta \tilde{N}_{t+1}$ term partially offsets the $\Delta \tilde{S}_{t+1}$ term. Strikingly, there is no shift in the Beveridge curve under constant matching efficiency.  This confirms the results from the two state model (and much of the literature) that the decline in matching efficiency was an important contributor to the loop in the Beveridge curve.

\section{Conclusion}\label{sec:conclusion}

The empirical Beveridge curve is easy to calculate, as it only requires data on the stocks of unemployed workers and job openings.  This ease of measurement may help explain the attention it has received.  Unfortunately, the Beveridge curve is (even in a simple model) the product of multiple factors, and can be difficult to interpret.  Our hope is that our results help clarify the behavior of the Beveridge curve and reconcile some conflicting ideas in the literature.

 We have shown that reduced-form matching efficiency, changes in the separation probability, and out-of-steady-state dynamics all played important roles in the recent shift of the Beveridge curve.  Comparing the pre-2010 period to the post-2010 period, out-of-steady-state dynamics and a fall in matching efficiency both pushed the curve upward, while the changes in the separation probability pushed the curve downward.  The net effect was the observed upward shift in the empirical Beveridge curve.  Our results are largely unchanged when we include a nonparticipation margin.  One area for more research is the effect of on-the-job search, which would affect the measurement of matching efficiency.     
 
A realistic model of the Great Recession therefore needs, (1) a mechanism for reduced-form matching efficiency to fall during and after the recession, (2) a non-constant separation probability, which can generate an increase in job losses towards the end of the recession.  Furthermore, models should not be evaluated using steady-state approximations, since the rapid changes in the labor market around the Great Recession made out-of-steady-state dynamics a first-order issue.

We reach similar conclusions regarding earlier recessions, though the role of matching efficiency is generally smaller.  Importantly, the relatively small Beveridge curve loops in earlier recessions were the product of changes in the separation probability nearly offsetting out-of-steady-state dynamics.  We find that these shifters move the intercept of the Beveridge curve continuously, not just at business cycle peaks and troughs.  As a result, the slope of the empirical Beveridge curve is distinct from the slope of the implied (constant separation probability, constant matching efficiency) steady-state curve.

\newpage

\bibliography{reference}

@ARTICLE{ahn2016,
author={Hie Joo Ahn and James D. Hamilton},
title={{Heterogeneity and Unemployment Dynamics}},
year = {2019},
journal = {Journal of Business and Economic Statistics},
volume = { },
number = { },
pages = {},
url = { DOI: 10.1080/07350015.2018.1530116}
}

@ARTICLE{barnichon2010,
title = {Building a composite Help-Wanted Index},
author = {Barnichon, Regis},
year = {2010},
journal = {Economics Letters},
volume = {109},
number = {3},
pages = {175-178},
url = {https://EconPapers.repec.org/RePEc:eee:ecolet:v:109:y:2010:i:3:p:175-178}
}

@TechReport{barnichon2010b,
  author={Regis Barnichon and Andrew Figura},
  title={{What drives movements in the unemployment rate? a decomposition of the Beveridge curve}},
  year=2010,
  month=9,
  institution={Board of Governors of the Federal Reserve System (U.S.)},
  type={Finance and Economics Discussion Series},
  url={https://ideas.repec.org/p/fip/fedgfe/2010-48.html},
  number={2010-48},
  doi={},
}

@article{barnichon2012,
title = "Which industries are shifting the Beveridge curve?",
author = "Regis Barnichon and Michael Elsby and Bart Hobijn and Aysegul Sahin",
year = "2012",
month = "6",
language = "English (US)",
volume = "135",
pages = "25--37",
journal = "Monthly Labor Review",
issn = "0098-1818",
publisher = "US Bureau of Labor Statistics",
number = "6",
}

@article{barnichon2015,
Author = {Barnichon, Regis and Figura, Andrew},
Title = {Labor Market Heterogeneity and the Aggregate Matching Function},
Journal = {American Economic Journal: Macroeconomics},
Volume = {7},
Number = {4},
Year = {2015},
Month = {October},
Pages = {222-49},
DOI = {10.1257/mac.20140116},
URL = {http://www.aeaweb.org/articles?id=10.1257/mac.20140116}}

@article{CET2015,
Author = {Christiano, Lawrence J. and Eichenbaum, Martin S. and Trabandt, Mathias},
Title = {Understanding the Great Recession},
Journal = {American Economic Journal: Macroeconomics},
Volume = {7},
Number = {1},
Year = {2015},
Month = {January},
Pages = {110-67},
DOI = {10.1257/mac.20140104},
URL = {http://www.aeaweb.org/articles?id=10.1257/mac.20140104}
}

@TechReport{daly2011,
  author={Daly, Mary C. and Hobijn, Bart and Valletta, Robert G.},
  title={{The Recent Evolution of the Natural Rate of Unemployment}},
  year=2011,
  month=Jul,
  institution={Institute for the Study of Labor (IZA)},
  type={IZA Discussion Papers},
  url={https://ideas.repec.org/p/iza/izadps/dp5832.html},
  number={5832},
  doi={},
}

@Article{davis2013,
  author={Steven J. Davis and R. Jason Faberman and John C. Haltiwanger},
  title={{The Establishment-Level Behavior of Vacancies and Hiring}},
  journal={The Quarterly Journal of Economics},
  year=2013,
  volume={128},
  number={2},
  pages={581-622},
  month={},
  doi={},
  url={https://ideas.repec.org/a/oup/qjecon/v128y2013i2p581-622.html}
}

@Article{diamond2015,
  author={Diamond, Peter A. and Sahin, Aysegul},
  title={{Shifts in the Beveridge curve}},
  journal={Research in Economics},
  year=2015,
  volume={69},
  number={1},
  pages={18-25},
  month={},
  doi={10.1016/j.rie.2015.01.002},
  url={https://ideas.repec.org/a/eee/reecon/v69y2015i1p18-25.html}
}

@article{eichenbaum2015,
author = {Martin S. Eichenbaum},
title = {Comment},
journal = {NBER Macroeconomics Annual},
volume = {29},
number = {1},
pages = {129-145},
year = {2015},
doi = {10.1086/680585},
URL = { https://doi.org/10.1086/680585  },
eprint = {https://doi.org/10.1086/680585}
}

@article{elsby2009,
Author = {Elsby, Michael W. L. and Michaels, Ryan and Solon, Gary},
Title = {The Ins and Outs of Cyclical Unemployment},
Journal = {American Economic Journal: Macroeconomics},
Volume = {1},
Number = {1},
Year = {2009},
Month = {January},
Pages = {84-110},
DOI = {10.1257/mac.1.1.84},
URL = {http://www.aeaweb.org/articles?id=10.1257/mac.1.1.84}}

@Article{elsby2010,
  author={Michael W. L. Elsby and Bart Hobijn and Aysegul Sahin},
  title={{The Labor Market in the Great Recession}},
  journal={Brookings Papers on Economic Activity},
  year=2010,
  volume={41},
  number={1 (Spring},
  pages={1-69},
  month={},
  doi={},
  url={https://ideas.repec.org/a/bin/bpeajo/v41y2010i2010-01p1-69.html}
}

@article{elsby2015,
Author = {Elsby, Michael W. L. and Michaels, Ryan and Ratner, David},
Title = {The Beveridge Curve: A Survey},
Journal = {Journal of Economic Literature},
Volume = {53},
Number = {3},
Year = {2015},
Month = {September},
Pages = {571-630},
DOI = {10.1257/jel.53.3.571},
URL = {http://www.aeaweb.org/articles?id=10.1257/jel.53.3.571}}

@Article{fujita2009,
  author={Shigeru Fujita and Garey Ramey},
  title={{The Cyclicality Of Separation And Job Finding Rates}},
  journal={International Economic Review},
  year=2009,
  volume={50},
  number={2},
  pages={415-430},
  month={May},
  doi={},
  url={https://ideas.repec.org/a/ier/iecrev/v50y2009i2p415-430.html}
}

@Article{furlanetto2016,
  author={Francesco Furlanetto and Nicolas Groshenny},
  title={{Mismatch Shocks and Unemployment During the Great Recession}},
  journal={Journal of Applied Econometrics},
  year=2016,
  volume={31},
  number={7},
  pages={1197-1214},
  month={November},
  doi={},
  url={https://ideas.repec.org/a/wly/japmet/v31y2016i7p1197-1214.html}
}

@article{hall_wohl2018,
Author = {Hall, Robert E. and Schulhofer-Wohl, Sam},
Title = {Measuring Job-Finding Rates and Matching Efficiency with Heterogeneous Job-Seekers},
Journal = {American Economic Journal: Macroeconomics},
Volume = {10},
Number = {1},
Year = {2018},
Month = {January},
Pages = {1-32},
DOI = {10.1257/mac.20170061},
URL = {http://www.aeaweb.org/articles?id=10.1257/mac.20170061}
}

@article{kroft2015,
author = {Kory Kroft and Fabian Lange and Matthew J. Notowidigdo and Lawrence F. Katz},
title = {Long-Term Unemployment and the Great Recession: The Role of Composition, Duration Dependence, and Nonparticipation},
journal = {Journal of Labor Economics},
volume = {34},
number = {S1},
pages = {S7-S54},
year = {2016},
doi = {10.1086/682390},
URL = {         https://doi.org/10.1086/682390},
eprint = { https://doi.org/10.1086/682390}
}

@Book{pissarides2000,
  author={Christopher A. Pissarides},
  title={{Equilibrium Unemployment Theory, 2nd Edition}},
  publisher={The MIT Press},
  year=2000,
  month={January},
  volume={1},
  series={MIT Press Books},
  edition={},
  doi={},
  isbn={ARRAY(0x56bedea0)},
  url={https://ideas.repec.org/b/mtp/titles/0262161877.html}
}

@techreport{michaillat2019,
 title = "Beveridgean Unemployment Gap",
 author = "Michaillat, Pascal and Saez, Emmanuel",
 institution = "National Bureau of Economic Research",
 type = "Working Paper",
 series = "Working Paper Series",
 number = "26474",
 year = "2019",
 month = "November",
 doi = {10.3386/w26474},
 URL = "http://www.nber.org/papers/w26474",
}

@Article{sahin2014,
  author={Aysegul Sahin and Joseph Song and Giorgio Topa and Giovanni L. Violante},
  title={{Mismatch Unemployment}},
  journal={American Economic Review},
  year=2014,
  volume={104},
  number={11},
  pages={3529-3564},
  month={November},
  doi={},
  url={https://ideas.repec.org/a/aea/aecrev/v104y2014i11p3529-64.html}
}

@Article{shimer2005,
  author={Robert Shimer},
  title={{The Cyclical Behavior of Equilibrium Unemployment and Vacancies}},
  journal={American Economic Review},
  year=2005,
  volume={95},
  number={1},
  pages={25-49},
  month={March},
  doi={},
  url={https://ideas.repec.org/a/aea/aecrev/v95y2005i1p25-49.html}
}

@Article{shimer2012,
  author={Robert Shimer},
  title={{Reassessing the Ins and Outs of Unemployment}},
  journal={Review of Economic Dynamics},
  year=2012,
  volume={15},
  number={2},
  pages={127-148},
  month={April},
  doi={10.1016/j.red.2011.01.002},
  url={https://ideas.repec.org/a/red/issued/11-293.html}
}

\newpage

\appendix 

\section{Full Decompositions}\label{sec:full_decompositions}
 
One may be concerned that results based on the Taylor approximation are not robust.  While the fit of the approximate Beveridge curve is strikingly good, it is not perfect.  Therefore, there is some room for non-linearities to affect the results.  A related issues is that the log-linearized Beveridge curve is not dynamically consistent:  If we plug implied vacancies into the matching function and the unemployment law of motion, we generally won't get the observed $U_{t+1}$ back.

In this section we decompose the shift in the empirical Beveridge curve using the exact vacancy equation rather than the log-linearized version.  Again, the goal is to measure the contributions to the shift due due to out-of-steady-state dynamics, changes in the separation probability, and changes in matching efficiency.

The starting point of our decomposition is the standard, steady-state Beveridge curve, with constant matching efficiency and separations:

\begin{equation}
V_{t}^{\overline{s},\overline{\sigma},\overline{\Delta U}}=\left[\frac{\overline{s}(1-U_{t})}{\overline{\sigma} U_{t}^{1-\alpha}}\right]^{1/\alpha}\label{eq:SS_bev}
\end{equation}

The steady state Beveridge curve sets $\Delta U_{t+1}=0$.  It  therefore the level of vacancies that would prevail after many months of constant $\overline{s}$ and $\overline{\sigma}$.

Let $t^{down}$ be a month from the downswing sample, and let $t^{up}$ be the corresponding (interpolated) period from the upswing with the same level of unemployment.  Then the observed vertical shift in the Beveridge curve is $V_{up}-V_{down}$.  The steady-state Beveridge curve \eqref{eq:SS_bev} obviously entails no shift, so $V_{up}^{\overline{s},\overline{\sigma},\overline{\Delta U}}-V^{\overline{s},\overline{\sigma},\overline{\Delta U}}_{down}=0$.  

We can define other counterfactual vacancy series.  We use superscripts with bars to denote that the margin is being held constant.  Thus, for example, 

\begin{align}
V_{t}^{\overline{\sigma} } & =\left[\frac{s_{t}(1-U_{t})- \Delta U_{t+1} }{\overline{\sigma}U_{t}^{1-\alpha}}\right]^{1/\alpha} \\
V_{t}^{\overline{\sigma},\overline{\Delta U}} & =\left[\frac{s_{t}(1-U_{t}) }{\overline{\sigma}U_{t}^{1-\alpha}}\right]^{1/\alpha}
\end{align}
with $V_{t}^{\overline{s}}$, $V_{t}^{\overline{s},\overline{\sigma}}$, $V_{t}^{\overline{s},\overline{\Delta U}}$,  and $V_{t}^{\overline{\Delta U}}$ defined similarly.

Next, consider the accounting identity

\begin{align}
V_{up}-V_{down} =&  \left( V_{up}-V_{down} \right)   -   \left( V_{up}^{\overline{\sigma}}-V^{\overline{\sigma}}_{down} \right) \nonumber \\
 & + \left( V_{up}^{\overline{\sigma}}-V^{\overline{\sigma}}_{down} \right)   -   \left( V_{up}^{\overline{s},\overline{\sigma}}-V^{\overline{s},\overline{\sigma}}_{down} \right) \nonumber \\
& + \left( V_{up}^{\overline{s},\overline{\sigma}}-V^{\overline{s},\overline{\sigma}}_{down} \right)   -  \left( V_{up}^{\overline{s},\overline{\sigma},\overline{\Delta U}}-V^{\overline{s},\overline{\sigma},\overline{\Delta U}}_{down} \right). \label{decomp}
\end{align}

This writes $V_{up}-V_{down}$ as three double differences.  The terms on the right hand side have the following interpretation:

\begin{itemize}
	\item $\left( V_{up}-V_{down} \right)   -   \left( V_{up}^{\overline{\sigma}}-V^{\overline{\sigma}}_{down} \right)$: The shift in the Beveridge curve accounted for by the time-variation in matching efficiency, conditional on having $s_{t}$ and $\Delta U_{t}$ at their observed values.    
    \item $\left( V_{up}^{\overline{\sigma}}-V^{\overline{\sigma}}_{down} \right)   -   \left( V_{up}^{\overline{s},\overline{\sigma}}-V^{\overline{s},\overline{\sigma}}_{down} \right) $: The shift accounted for by time-variation in the separation probability, conditional on having $\Delta U_{t}$ at its observed values and $\sigma$ held constant.
    \item $\left( V_{up}^{\overline{s},\overline{\sigma}}-V^{\overline{s},\overline{\sigma}}_{down} \right)   -  \left( V_{up}^{\overline{s},\overline{\sigma},\overline{\Delta U}}-V^{\overline{s},\overline{\sigma},\overline{\Delta U}}_{down} \right) $: The shift accounted for by time-variation in $ \Delta U_{t+1}$, conditional on having matching efficiency and the separation probability held constant.  Note that $V_{up}^{\overline{s},\overline{\sigma},\overline{\Delta U}}-V^{\overline{s},\overline{\sigma},\overline{\Delta U}}_{down} =0$ by construction.   
\end{itemize}

Thus, we can interpret equation \eqref{decomp} as moving us from the steady-state Beveridge curve (which cannot shift by construction) to the observed shift, by successively adding the observed time-variation in margins.  Equation \eqref{decomp} first adds observed dynamics, then adds observed the separation probability, then adds observed matching efficiency.  With three margins there are six possible orderings, and the results will, in general, depend on the ordering.

Table \ref{tab:nl_decomp_table} shows the results of all six orderings.  The results are remarkably consistent.  In all versions, separations push the Beveridge curve down during the upswing period, relative to the downswing period.  Both dynamics and matching efficiency have the opposite effect, contributing to the counter-clockwise loop in the observed Beveridge curve.  Generally, the contribution of matching efficiency is larger than that of dynamics, sometimes dramatically so. The only outlier is the fourth row.  However, we believe that the first two rows are the most important, because they put $\Delta U_{t+1}$ first in the ordering, which ensures dynamic consistency.  

\begin{table}
\centering
\caption{Contributions to the Shift in the Beveridge Curve}
\label{tab:nl_decomp_table}
\begin{threeparttable}
\sisetup{                                                            
  input-symbols=(),                                                  
  table-format=-1.5,                                                 
  table-space-text-post=***,                                         
  table-align-text-post=false,                                       
  group-digits=false                                                 
}                                                                    
\sisetup{                                                            
  input-symbols=(),                                                  
  table-format=-1.5,                                                 
  table-space-text-post=***,                                         
  table-align-text-post=false,                                       
  group-digits=false                                                 
}                                                                    
\begin{tabular}{S|SSS}                                               
{Ordering} & {Dynamics} & {Separations} & {Matching} \\ \hline \hline
{$\Delta U_{t+1}$, $s$, $\sigma$} & 115.26 &-177.18 & 161.92\\       
{$\Delta U_{t+1}$, $\sigma$, $s$} & 115.26 &-444.82 & 429.55\\       
{$s$, $\Delta U_{t+1}$, $\sigma$} & 121.46 &-183.38 & 161.92\\       
{$s$, $\sigma$, $\Delta U_{t+1}$} & 213.19 &-183.38 & 70.19\\        
{$\sigma$, $\Delta U_{t+1}$, $s$} & 205.80 &-444.82 & 339.01\\       
{$\sigma$, $s$, $\Delta U_{t+1}$} & 213.19 &-452.21 & 339.01\\       
\end{tabular}                                                        

  \begin{tablenotes}
\item \footnotesize
\textit{Notes}: Percentage point contributions to the vertical shift in the Beveridge curve, averaged over the ``downswing'' sample points discussed earlier.  ``Ordering'' column shows the order in which margins are set to their observed values.  For example, the row $\Delta U_{t+1},s,\sigma$ starts with the steady-state curve, then adds observed $\Delta U_{t+1}$, then adds observed $s_{t}$, and finally adds the observed $\sigma_{t}$.    
\end{tablenotes}
\end{threeparttable}
\end{table}

Nearly all of the contributions in Table \ref{tab:nl_decomp_table} are well above 100 percent.  This shows just how important all three margins are in understanding the shift of the Beveridge curve.  The shift we observe empirically is relatively small, when compared to the effects of the shifters taken separately. 



\end{document}